\documentclass[aps,prx,reprint,preprintnumbers,superscriptaddress,nofootinbib,longbibliography,floatfix]{revtex4-2}
\pdfoutput=1
\usepackage{rotating}
\usepackage{array}
\usepackage{amsmath}
\usepackage[normalem]{ulem}
\usepackage{slashed}
\usepackage{booktabs}
\usepackage[pdftex,table]{xcolor}
\usepackage{units}
\usepackage{xfrac}
\usepackage{mathtools}
\usepackage{empheq}
\usepackage[]{units}
\usepackage{multirow}
\usepackage{amssymb}
\usepackage{url}
\usepackage{comment}
\usepackage{physics}
\usepackage{color,soul}
\usepackage{bbm}
\usepackage[caption=false]{subfig}

\usepackage{hyperref}
\hypersetup{
  colorlinks=true,
  citecolor=blue,
  linkcolor=blue,
  urlcolor=blue
}

\newcommand{\geant}{\textsc{Geant}}
\newcommand{\calosc}{\textsc{CaloScore}}
\newcommand{\calosctwo}{\textsc{CaloScore}~v2}
\newcommand{\xbf}{\textbf{x}}

\begin{document}

\title{CaloScore v2: Single-shot Calorimeter Shower Simulation with Diffusion Models}

\author{Vinicius Mikuni}
\email{vmikuni@lbl.gov}
\affiliation{National Energy Research Scientific Computing Center, Berkeley Lab, Berkeley, CA 94720, USA}

\author{Benjamin Nachman}
\email{bpnachman@lbl.gov}
\affiliation{Physics Division, Lawrence Berkeley National Laboratory, Berkeley, CA 94720, USA}
\affiliation{Berkeley Institute for Data Science, University of California, Berkeley, CA 94720, USA}

\begin{abstract}
    Diffusion generative models are promising alternatives for fast surrogate models, producing high-fidelity physics simulations. However, the generation time often requires an expensive denoising process with hundreds of function evaluations, restricting the current applicability of these models in a realistic setting. In this work, we report updates on the \calosc~architecture, detailing the changes in the diffusion process, which produces higher quality samples, and the use of progressive distillation, resulting in a diffusion model capable of generating new samples with a single function evaluation. We demonstrate these improvements using the Calorimeter Simulation Challenge 2022 dataset.
\end{abstract}

\maketitle


\section{Introduction}
\label{sec:intro}

Deep generative models are a disruptive technology, enhancing many aspects of every day life and basic science research.  In high energy physics, calorimeter simulations have been a benchmark for new deep generative models since their first application.  Detailed physics simulations of particle showers in calorimeters are often prohibitively slow due to the large number of secondary particles produced as the primary particle is stopped inside the detector material.  Bespoke and often proprietary fast simulations have been developed for many cases, but they are usually derived using low-dimensional heuristics.  Deep learning has the potential to match the quality of detailed simulations in their full high-dimensional representation while also matching the speed of classical fast simulations.  Automated, high-fidelity and fast calorimeter simulations can enhance the science of existing detectors and catalyze the development of better detectors at future experiments.

\nocite{GANs} 
The application of deep generative models to calorimeter simulation began with CaloGAN~\cite{GanPhys2,GanPhys3}.  Since that time, Generative Adversarial Networks (GANs)~\cite{GANs}~\cite{GanPhys2,GanPhys3,deOliveira:2017rwa,Erdmann:2018kuh,Erdmann:2018jxd,Belayneh:2019vyx,Vallecorsa:2019ked,SHiP:2019gcl,Chekalina:2018hxi,Carminati:2018khv,Vallecorsa:2018zco,Musella:2018rdi,Deja:2019vcv,ATLAS:2022jhk,ATL-SOFT-PUB-2018-001,ATLAS:2021pzo}, Variational Autoencoders~\cite{VAEs}~\cite{ATL-SOFT-PUB-2018-001,ATLAS:2022jhk,Buhmann:2021lxj,Buhmann:2021caf,Diefenbacher:2023prl}, Normalizing Flows (NFs)~\cite{NFs}~\cite{caloflow1,caloflow2,Buckley:2023rez,Krause:2022jna,Diefenbacher:2023vsw,Cresswell:2022tof, Liu:2023lnn}, and Diffusion Models~\cite{scoremodels}~\cite{mikuni:caloscore,Buhmann:2023bwk,Acosta:2023zik} have been applied to this problem.  This research entered a precision era with the first NF application (CaloFlow)~\cite{caloflow1}, which showed that even a post-hoc classifier had difficulty distinguishing physics from machine learning simulators.  A number of related innovations in the CaloFlow series are motivating for our work including teacher-student training~\cite{caloflow2} and factorizing into energy/layer and shape/layer.
  Deep generative models are also now being used in practice.  The ATLAS experiment has integrated a GAN into its fast simulation, which has improved the modeling of hadronic final states~\cite{ATLAS:2021pzo}.  Simulations from ATLAS, combined with additional samples from more granular hypothetical detectors, form the CaloChallenge~\cite{calochallenge}, a data challenge to compare a diverse set of models on the same calorimeter simulations.  The score-based model CaloScore~\cite{mikuni:caloscore} was the first model deployed on all three datasets from the CaloChallenge.  Since that time, NF-based approaches have also been studied for CaloChallenge datasets 1~\cite{Krause:2022jna,Cresswell:2022tof} and 2-3~\cite{Buckley:2023rez}.  As the performance of these approaches has not been quantified with exactly the same metric\footnote{A forthcoming CaloChallenge review paper will do this carefully, also including many methods that have not (yet) been published as standalone papers.}, it is hard to know which is `best', but it is clear that they are all able to accurately describe various aspects of the complex calorimeter showers.

Since the publication of \calosc, we have improved the performance significantly by introducing a number of innovations.  Collectively, these updates constitute \calosctwo, which represents the state of the art in calorimeter emulation.  Improvements to the architecture and training procedure result in a model that has significantly better fidelity and is much faster than the original \calosc.  One aspect of \calosctwo~is progressive distillation to reduce the number of timesteps in the diffusion process, with one step already achieving reasonable fidelity.  Additionally, we modify the diffusion process to decrease the loss variance during training, and separate the task of determining the total energy deposition with the voxel generation through an additional generative model. Altogether, \calosctwo~is essentially a new model built on the foundation of the original \calosc~-- where we demonstrated that diffusion models are a compromise between flexibility (easy for GANs, hard for NFs) and robustness (easy for NFs, hard for GANs) -- with qualitatively superior performance than its predecessor.

This paper is organized as follows.  Section~\ref{sec:score} introduces Diffusion Models and Sec.~\ref{sec:sampling} describes how we sample from a trained model.  The three CaloChallenge datasets are detailed in Sec.~\ref{sec:dataset}.  The properties of \calosctwo~are provided in Sec.~\ref{sec:model} before presenting numerical results in Sec.~\ref{sec:results}.  The paper ends with conclusions and outlook in Sec.~\ref{sec:conclusions}.

\section{Diffusion Models}
\label{sec:score}
Diffusion generative models apply perturbations to the data to slowly corrupt the initial dataset into a tractable noise distribution. The generation step aims to reverse this processes, starting from a noise distribution that is denoised towards realistic examples of the data to be generated. The time-dependent perturbation can be described by the following stochastic differential equation (SDE):
\begin{equation}
    \mathrm{d}\xbf = f(\xbf,t)\mathrm{d}t + g(t)\mathrm{dw}.
\end{equation}
In this equation, the data $\xbf \in \mathbb{R}^{d}$ are perturbed over a time parameter $t \in [0,1]$ with perturbation parameters defined by the choice of drift and diffusion coefficients $f(\xbf,t) \in \mathbb{R}^d$ and $g(t) \in \mathbb{R}$, respectively. The stochastic term is identified by the Wiener process, or Brownian motion,  $w(t)\in\mathbb{R}^d$, often sampled from a normal distribution with the same dimension as the data. To reverse this processes towards the generation of new data, the reverse stochastic differential equation needs to be solved, described by
\begin{equation}
    \mathrm{d}\xbf = [f(\xbf,t)-g(t)^2\nabla_x\log p(\xbf)]\mathrm{d}t + g(t)\mathrm{d\bar{w}}.
    \label{eq:rsde}
\end{equation}

While the forward SDE is easy to solve, the reverse process requires the knowledge of the term $\nabla_x\log p(\xbf)$, also known as the score function of the data. Since $\xbf$ is high-dimensional, the probability density of the data $p(\xbf)$ is often intractable, and similarly, the score function cannot be easily estimated. Alternatively, the authors in~\cite{Song2021ScoreBasedGM} have shown that, in the limit of small noise perturbations, learning the score function of perturbed data is equivalent to learning the score function of the data itself. This observation motivates the loss function 
\begin{equation}
    \mathcal{L} = \frac{1}{2}\mathbb{E}_{\xbf_t,t}\left [ \lambda(t)\left \| s_\theta(\xbf_t,t) -  \nabla_{\xbf_t} \log q(\xbf_t|\xbf)\right\| ^2_2\right ].
    \label{eq:loss_score_time}
\end{equation}

The neural network $s_\theta(\xbf_t,t)$ with trainable parameters $\theta$ takes as input data $\xbf_t$ that has been perturbed at time $t$. The weight parameter $\lambda(t)$ is a positive function used to determine  the importance of each term in the loss function over time. By considering Gaussian perturbation $q(\xbf_t|\xbf) = \mathcal{N}(\xbf_t,\alpha_t\xbf,\sigma_t^2\textbf{I})$,  the score function of the perturbed data $\xbf_t = \alpha_t\xbf + \sigma_t\epsilon$, $\epsilon\sim\mathcal{N}(0,1)^d$ is identified as:
\begin{equation}
     \nabla_{\xbf_t} \log q(\xbf_t|x) = \frac{\xbf-\xbf_t}{\sigma_t^2} \sim \frac{\mathcal{N}(0,1)^d}{\sigma_t}.
     \label{eq:noise_score}
\end{equation}

In the original \calosc~implementation, $\lambda(t) \equiv \sigma_t^2$, which improves the training stability by removing the $\sigma$-dependence of the perturbed score function in Eq.~\ref{eq:noise_score}. While the direct prediction of the score function is beneficial, recent works have moved towards learning different representations of Eq.~\ref{eq:loss_score_time}. The reason for this change is explained by the high variance of the signal-to-noise-ratio (SNR) distribution $\alpha_t/\sigma_t$. At the beginning the diffusion process, at time values near zero, the standard deviation of the perturbation is designed to be small, leading to large values of SNR. Conversely, at time values near one, the perturbation is the largest to ensure that any prior data distribution is diffused towards a normal distribution at the end of the diffusion process, leading to small values of SNR. Since we expect $\sigma_ts_\theta(\xbf_t,t)\sim\mathcal{N}(0,1)^d$, the expected values of $s_\theta(\xbf_t,t)$ also show high variance, requiring $s_\theta(\xbf_t,t)$ to spam a wide range of values. In \calosctwo, we instead opt for a so-called velocity implementation, introduced in~\cite{salimans2022progressive} that defines a target $\mathbf{v}_t \equiv \alpha_t\mathbf{\epsilon}-\sigma_t\mathbf{x}$ which modifies the loss function in Eq.~\ref{eq:loss_score_time} to introduce the updated loss as
\begin{equation}
    \mathcal{L} = \mathbb{E}_{\xbf_t,t} \left\| \mathbf{v}_t - \mathbf{v}_{\theta}(\xbf_t,t)\right\|^2,
    \label{eq:loss_v2}
\end{equation}
with a neural network trained to learn directly the velocity parameter while taking as inputs the time and the perturbed data. From this implementation, we can still identify the approximation of the score function of the perturbed data as
\begin{equation}
    s_\theta(\xbf_t,t) =  \xbf_t  - \frac{\alpha_t}{\sigma_t}\mathbf{v}_\theta(\xbf_t,t),
\end{equation}
with the advantage of having the velocity parameter with similar range over the entire time interval of the diffusion process. 

The choice of the drift and diffusion coefficients $f(\xbf,t)$ and $g(t)$ are also important parameters of the diffusion process. In \calosc, different choices of parameters were investigated yielding similar performance. The parameters of the perturbation $\alpha$ and $\sigma$ can also be used to define $f(\xbf,t)$ and $g(t)$ with
\begin{equation}
    \begin{split}  
        f(\xbf,t) &= \frac{\mathrm{d}\log \alpha_t}{\mathrm{dt}}\xbf_t \\
        g^2(t) &= \frac{\mathrm{d}\sigma^2_t}{\mathrm{dt}} -2 \frac{\mathrm{d}\log \alpha_t}{\mathrm{dt}} \sigma^2_t.
    \end{split}
    \label{eq:fg}
\end{equation}
For \calosctwo, we choose to focus on the variance preserving (VP) implementation which additionally requires $\sigma^2_t + \alpha^2_t =1$. A cosine schedule is used with $\alpha_t =\cos(0.5\pi t)$ and $\sigma_t=\sin(0.5\pi t)$. This choice is in contrast with the previous $\beta$-parameterization used in \calosc~, where $f(\xbf,t) = -\frac{1}{2}\beta(t)\xbf$ and $g(t) = \sqrt{\beta(t)}$ with $\beta(t) = \beta_{\textrm{min}} + t \left( \beta_{\textrm{max}} - \beta_{\textrm{min}} \right)$ with $\beta_{\textrm{min}} = 0.1$ and $\beta_{\textrm{max}}=20$. This update is motivated by the use of the progressive distillation method, explained in further detail in Section~\ref{sec:sampling}.

\section{Sample Generation}
\label{sec:sampling}
With the approximation of the score function of the data, different methods an be employed to generate new observations. Stochastic solvers can be used to solve Eq.~\ref{eq:rsde}. In \calosc, sampling is performed using the Euler-Maruyama algorithm~\cite{kloeden1992stochastic} followed by an additional corrector step that uses the Langevin MCMC approach~\cite{Parisi:1980nh,Grenander1994REPRESENTATIONSOK} to increase the sampling quality. This approach, however requires the discretization over time of the reverse SDE in Eq.~\ref{eq:rsde} to be of $\mathcal{O}(100)$ to generate high fidelity calorimeter images. The large number of discretization steps is a natural consequence of the stochastic nature of the equation to be solved, since the precision in this case is determined by the magnitude of the stochastic noise added in each step. On the other hand, the reverse SDE admits a deterministic solution~\cite{Song2021ScoreBasedGM} of the form
\begin{equation}
    \frac{\mathrm{d}\xbf_t}{\mathrm{d}t} = f(\xbf,t)-\frac{1}{2}g(t)^2\nabla_x\log q(\xbf_t)\,,
    \label{eq:ode}
\end{equation}
which can be solved with fewer time steps, while also providing an unique mapping between points of the initial noise distribution and the generated data. While Eq.~\ref{eq:ode} can be solved as is with direct integration, the authors of Ref.~\cite{DBLP:journals/corr/abs-2010-02502} propose a different deterministic sampler named DDIM, also shown to represent an integration rule for Eq.~\ref{eq:ode}, but requiring fewer time steps to achieve the same level of precision. In the DDIM solver, the update rule is then specified by:
\begin{equation}
    \mathbf{x}_s = \alpha_s\mathbf{x}_\theta(\mathbf{x}_t,t)  + \sigma_s\frac{\mathbf{x}_t -\alpha_t\mathbf{x}_\theta(\mathbf{x}_t,t)}{\sigma_t},
\end{equation}
for time $s<t$ and  position prediction $\mathbf{x}_\theta(\mathbf{x}_t,t) = \alpha_t\xbf_t - \sigma_t\mathbf{v}_\theta(\xbf_t,t)$. Additionally, the choice of the DDIM sampler is also motivated by the use of progressive distillation~\cite{salimans2022progressive}. The idea of progressive distillation is to introduce a second model whose task is to learn to halve the number of time steps required by the DDIM solver using a trained diffusion model as a guide. In this approach, the trained diffusion model (``teacher'') is used to initialize a ``student'' model. During training, the goal of the student model is to denoise data $\mathbf{x}_t$ towards a target $\mathbf{\widetilde{x}}_t$. The difference is that  $\mathbf{\widetilde{x}}_t$ does not represent the clean data ($\mathbf{x}$), but is instead one that makes a single student DDIM step to match two teacher DDIM steps. After the student model is trained, generation can be performed using half the number of time steps compared to the teacher model. This process is then repeated multiple times, with the student at the end of each iteration becoming the new teacher. In \calosctwo, the initial diffusion model uses 512 time steps to ensure precision and is distilled multiple times with results using 8 time steps and a single time step reported.  

\section{Fast Calorimeter Simulation Challenge 2022}
\label{sec:dataset}
The performance of \calosctwo~is evaluated using the datasets released for the Fast Calorimeter Simulation Challenge 2022~\cite{faucci_giannelli_michele_2022_6366324,michele_faucci_giannelli_2022_6368338,faucci_giannelli_michele_2022_6366271,ATLAS:2021pzo}. Three datasets are provided, representing calorimeter shower simulations with \textsc{Geant}4~\cite{geant4} of different detector geometries and number of detector components. Dataset 1 \cite{michele_faucci_giannelli_2022_6368338} is based on the ATLAS  open dataset \cite{ATL-SOFT-PUB-2020-006,ATLAS:2021pzo} and is similar to the current ATLAS detector calorimeter geometry. While samples consisting of both photons and pions are provided, we evaluate our model  using only the photon dataset. The voxelization procedure is defined such that it reduces the amount of empty voxels, while maintaining high fidelity compared to the full simulation. The downside of this approach is that the geometrical information present in the original detector layout is also reduced since each voxel now covers a different area depending on the number of detector components merged during the voxelization. A total of 368 voxels are then left to describe the full detector slice. Photon energies are provided in this configuration for 15 incident energies ranging from 256 MeV up to 4 TeV in steps given by powers of two. For each generated energy, 10k samples are provided with this number reduced at higher energies due to long simulation times, resulting in a total of 121k used during training. 

Datasets 2 \cite{faucci_giannelli_michele_2022_6366271} and 3 \cite{faucci_giannelli_michele_2022_6366324} contain each 100k examples to be used for training  and are simulated using a common detector layout but with different voxelization granularity. The detector simulated has a concentric cylinder geometry with 45 layers, where each layer consists of active (silicon) and passive (tungsten) material, simulated with \textsc{Geant4}. Simulations for electrons are generated at the detector surface with initial energy sampled from a log-uniform distribution ranging from 1~GeV to 1~TeV. In dataset 2, each layer consists of 144 readout cells, with 9 in the radial and 16 in the angular directions. Dataset 3 is more granular, consisting of 900 readout cells in each layer, with 18 in the radial and 50 in the angular directions with a total of 6480 and 40500 voxels, respectively.

Since the initial representation of the datasets 2 and 3 are given in cylindrical coordinates, a preprocessing step was used in \calosc~to convert the datasets to Cartesian coordinates. This choice avoids the need for the generative model to learn the periodic boundary conditions in the $\alpha$ direction, while also centralizing the detector readouts. Unfortunately, this transformation is not reversible, since multiple voxels can be mapped to a single voxel in the new Cartesian representation\footnote{A one-to-one assignment between the two sets of coordinates is possible, but requires the distance interval in Cartesian coordinates to follow a non-linear function since the transformation of coordinates is itself non-linear.}. This choice limited the comparison of \calosc~ with other generative models and has now been abandoned.

Similarly, \calosctwo~uses a different data preprocessing to improve the fidelity of the generative model compared to \calosc. In the original \calosc, each voxel energy $E_v$ is normalized by the value of the  energy of the incident particle $E_0$ times a factor $f$ used to fix the energy scale normalization caused by the sampling fraction of the detector and ensure the normalized voxel energy $E'_v = \frac{E_v}{fE_0}$ lies between 0 and 1. In \calosctwo, we instead split the generation into two tasks: one that generates the overall deposited energy per layer of the calorimeter, and one that generates normalized voxel distributions. For the second task, the preprocessing is changed, with the normalization factor for each voxel to be the deposited energy per layer instead of the incident particle energy. For the first task, to determine the overall energy per layer, we first divide the deposited energy per layer by the initial particle energy, multiplied by the factor $f$. The additional preprocessing steps applied to the data are then identical to \calosc.  The normalized energy depositions are then transformed to logit-space, similarly to the strategy used in \textsc{CaloFlow}. The log-transformed value $u_v$ is defined as:
\begin{equation}
    u_v=\log \left ( \frac{x}{1-x} \right ), x=\alpha + (1-2\alpha)E'_v.
    \label{eq:log_transform}
\end{equation}
The value $\alpha$ in Eq.~\ref{eq:log_transform} is set to $10^{-6}$ and avoids a discontinuity when $E'_v=0$. The generated particle energy, used as a conditional input to the model, is also transformed before training. The transformed conditional energy $u_0$ is defined as:
\begin{equation}
    u_0 = \frac{e_0-e_{\textrm{min}}}{e_{\textrm{max}}-e_{\textrm{min}}},
\end{equation}
where $e_{\textrm{min}}$ and $e_{\textrm{max}}$ are the minimum and maximum energies available in the dataset used for the training. Last, all voxels and energy depositions per layer are standardized to have mean zero and unit variance across all training samples. 

\section{Model Architecture and Training Details}
\label{sec:model}
In the previous \calosc~implementation, the transformation to Cartesian coordinates resulted in a model that could be efficiently learned using few convolutional layers with large kernel sizes, implemented with a \textsc{U-net}~\cite{ronneberger2015u} network architecture.  In \calosctwo, we employ a similar \textsc{U-net} architecture, but include additional attention layers. More specifically, datasets 2 and 3 have the number of spatial components in each dimension reduced by a factor 2 every other convolutional layer (resulting in a factor $2\times2\times2=8$ reduction) with fixed kernel size set to 3. This process is repeated 3 times, with lowest dimensional representation reduced by a factor 512 compared to the initial number of voxels. The 3D convolution operations use 32, 64, and 96 hidden nodes  with swish~\cite{ramachandran2017searching} activation function. The attention layer is only used at the lowest dimensional representation, with data patches determined by the flattened array describing the data at the lowest dimensionality. The upsampling section of the architecture is a mirrored version, with dimensions increased by a factor 8 every other layer. Skip connections between the downsampling and upsampling sides of the architecture are combined with a concatenation operation, completing the architecture. Conditional information consisting of the time information, incident particle energy, and deposited energy per layer (in case of the diffusion model trained to generate normalized voxels), are included through an addition operation after every convolutional layer. A trainable embedding of the conditional features is created by a fully connected layer over the conditional inputs. The output size is fixed to match the expected output size of the convolutional layers. For dataset 1, the strategy is similar. The number of voxels to be simulated are reduced by a factor 2 every other layer, with this process repeated 4 times and overall reduction of factor 16 compared to the initial size. The number of hidden nodes for the 1D convolutional layers is then chosen to be 16, 32, 64, and 96 for each fixed dimensionality. Since this dataset is smaller compared to datasets 2 and 3, attention layers are used in all lower dimensional representations of the initial data.

A second diffusion model is introduced in \calosctwo, tasked to learn only the energy deposition per layer. The model used to train the diffusion model is based on the \textsc{ResNet}~\cite{he2016deep} architecture, consisting of multiple fully connected layers with additional skip connections. The number of \textsc{ResNet} layers is set to 3 in both datasets with 128 hidden nodes in dataset 1 and 1024 in datasets 2 and 3. Additional choices of hyperparameters such as overall number of layers and hidden node sizes were tested and did not yield noticeable improvements.

The training is carried out using the Perlmutter supercomputer interfaced with the Horovod package~\cite{sergeev2018horovod} for distributed training. 16 NVIDIA A100 GPUs are used simultaneously during training, while a single GPU is used for evaluation and timing comparison. All models are trained for up to 250 epochs with a cosine learning rate schedule~\cite{DBLP:journals/corr/LoshchilovH16a} with initial learning rate of $16\times10^{-4}$. If the loss function does not decrease for 30 consecutive epochs, evaluated in a separate testing set representing 20\% to the sample size, the training is stopped. The implementation of all models is carried out using \textsc{Keras} backend~\cite{keras} with \textsc{TensorFlow}~\cite{tensorflow}.

\section{Results}
\label{sec:results}

We evaluate the performance of \calosctwo~using the metrics available for the evaluation of the Fast Calorimeter Simulation Challenge 2022, as well as additional studies to quantify the agreement of different physics distributions with the original \geant~ simulations. Since dataset 3 before distillation is much slower than datasets 1 and 2, only distilled results with 8 and a single time step are reported.

\begin{figure*}[ht]
\centering
    \includegraphics[width=0.3\textwidth]{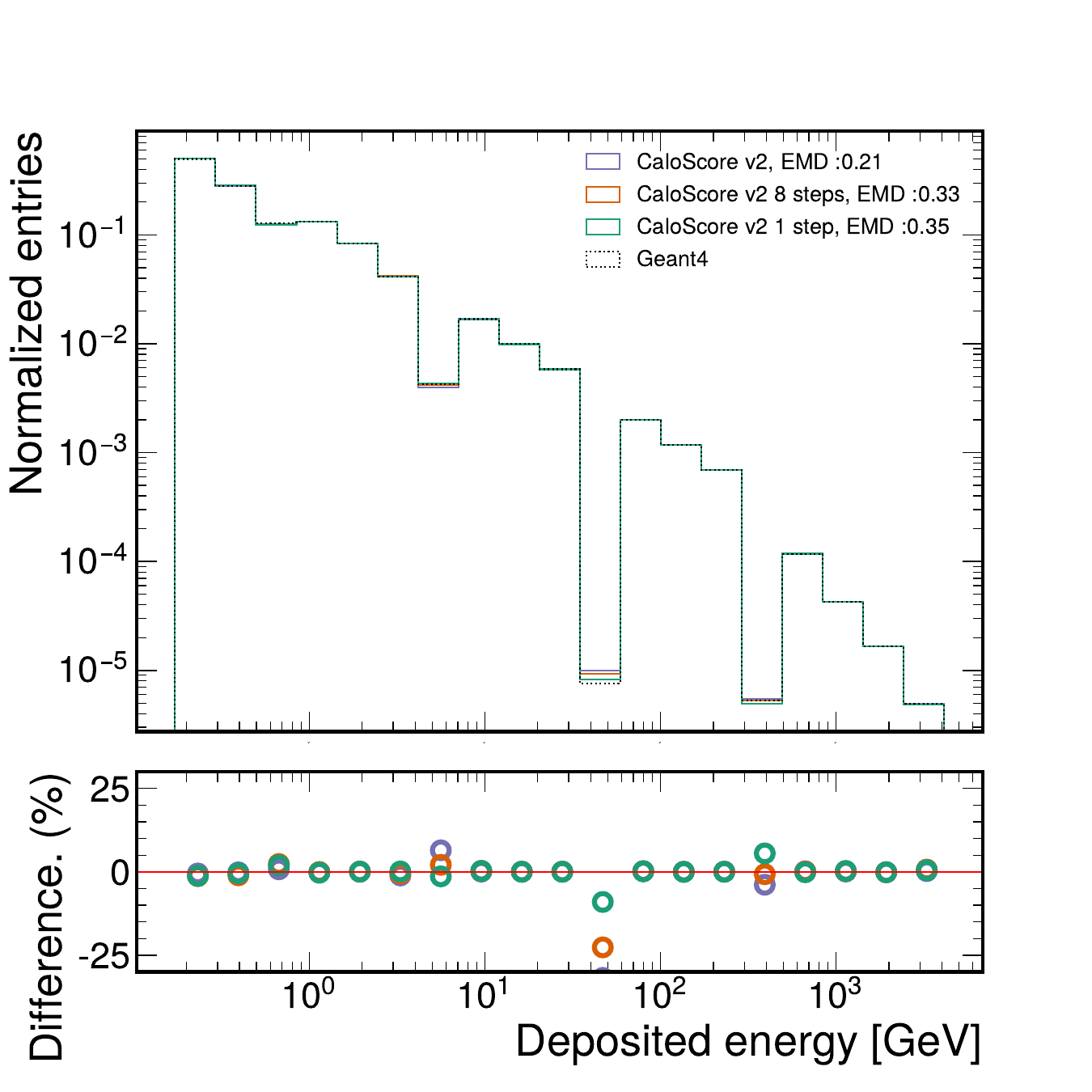}
    \includegraphics[width=0.3\textwidth]{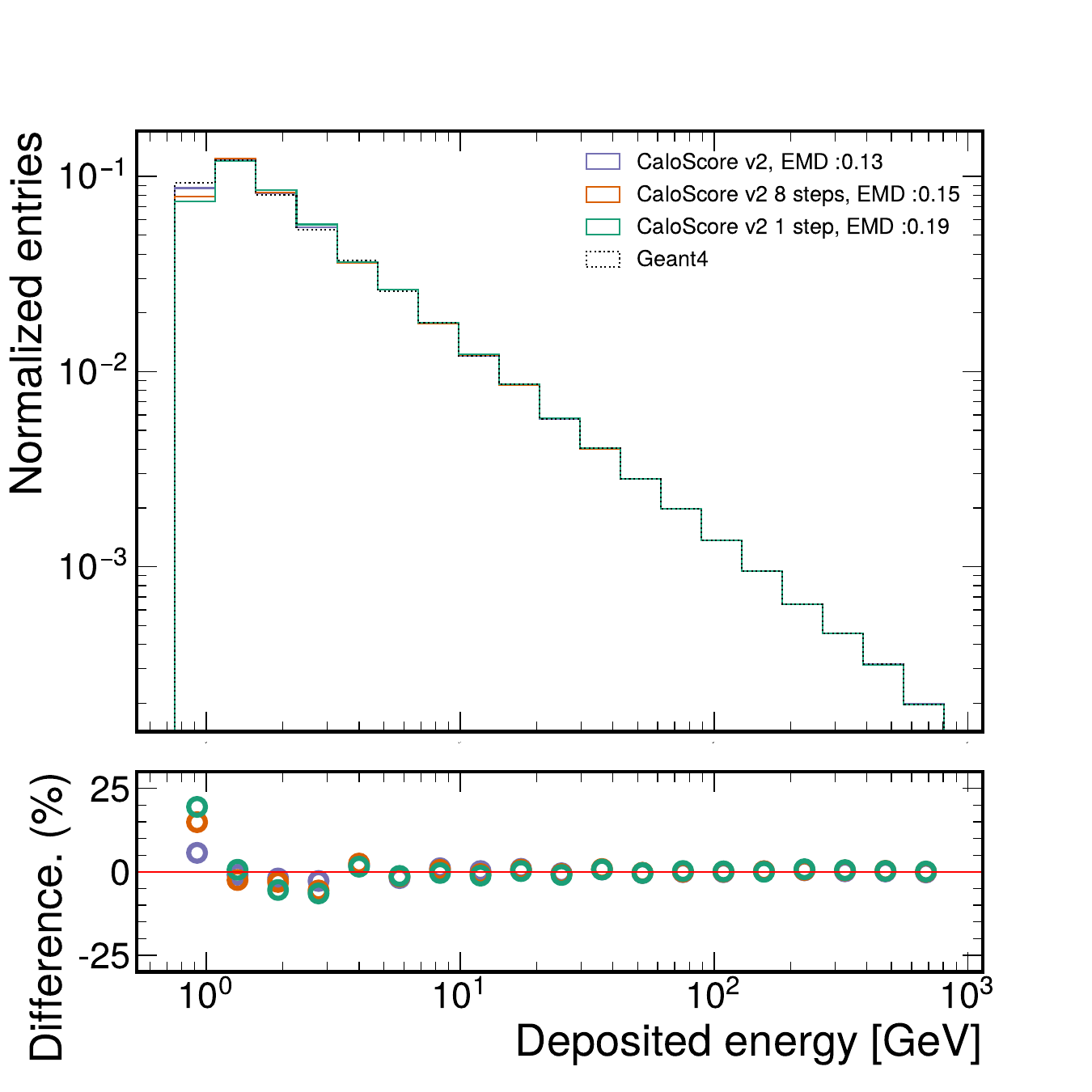}
    \includegraphics[width=0.3\textwidth]{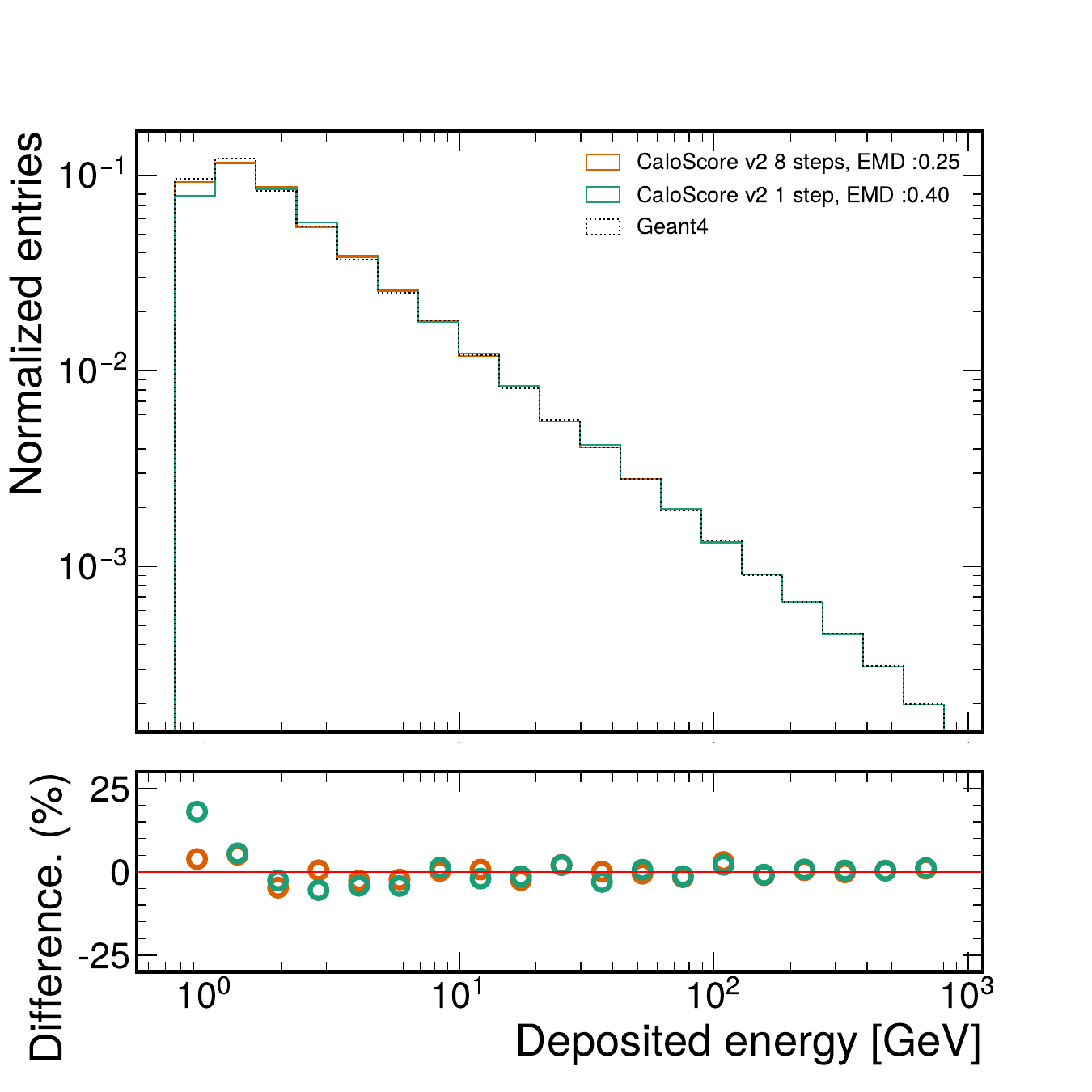}
    \includegraphics[width=0.3\textwidth]{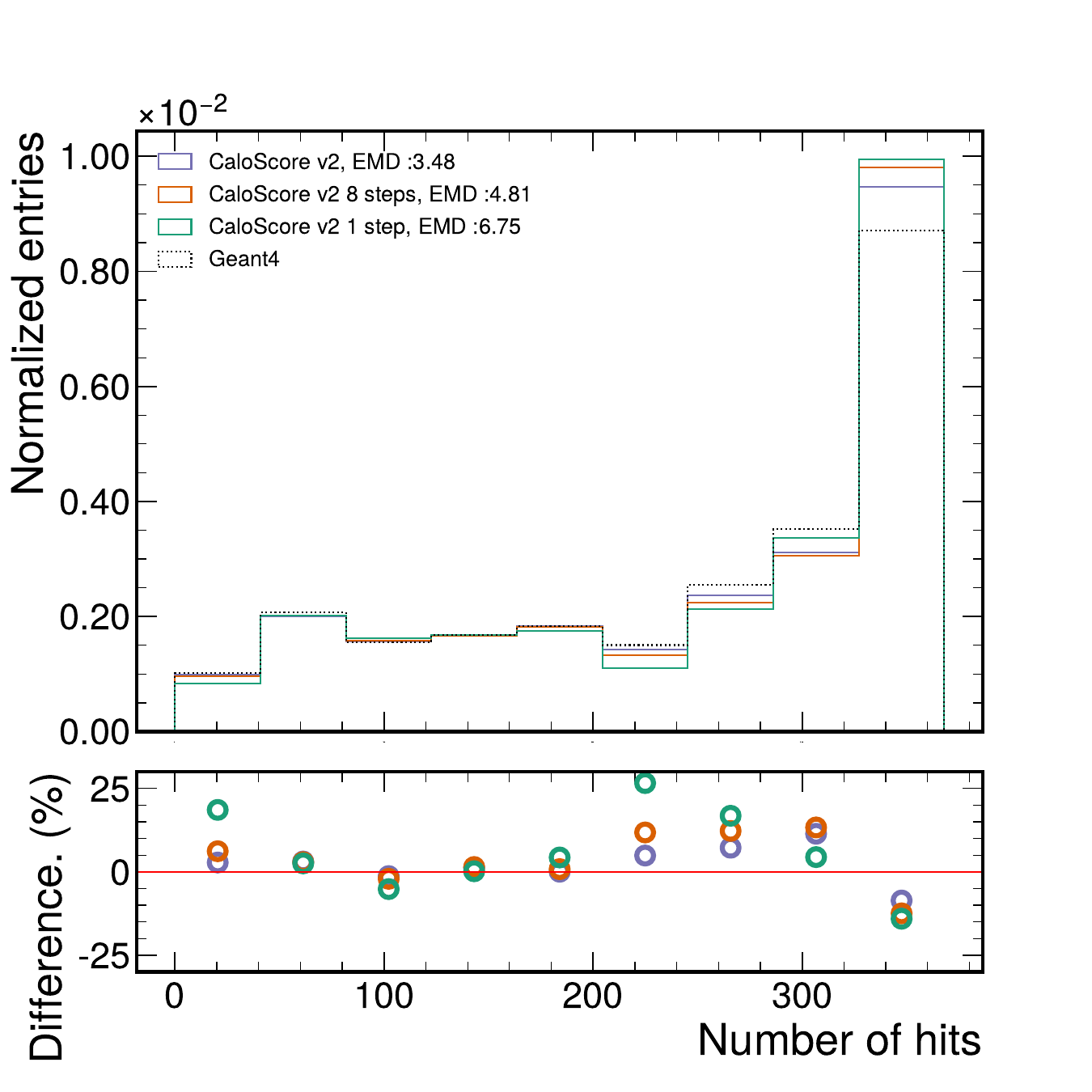}
    \includegraphics[width=0.3\textwidth]{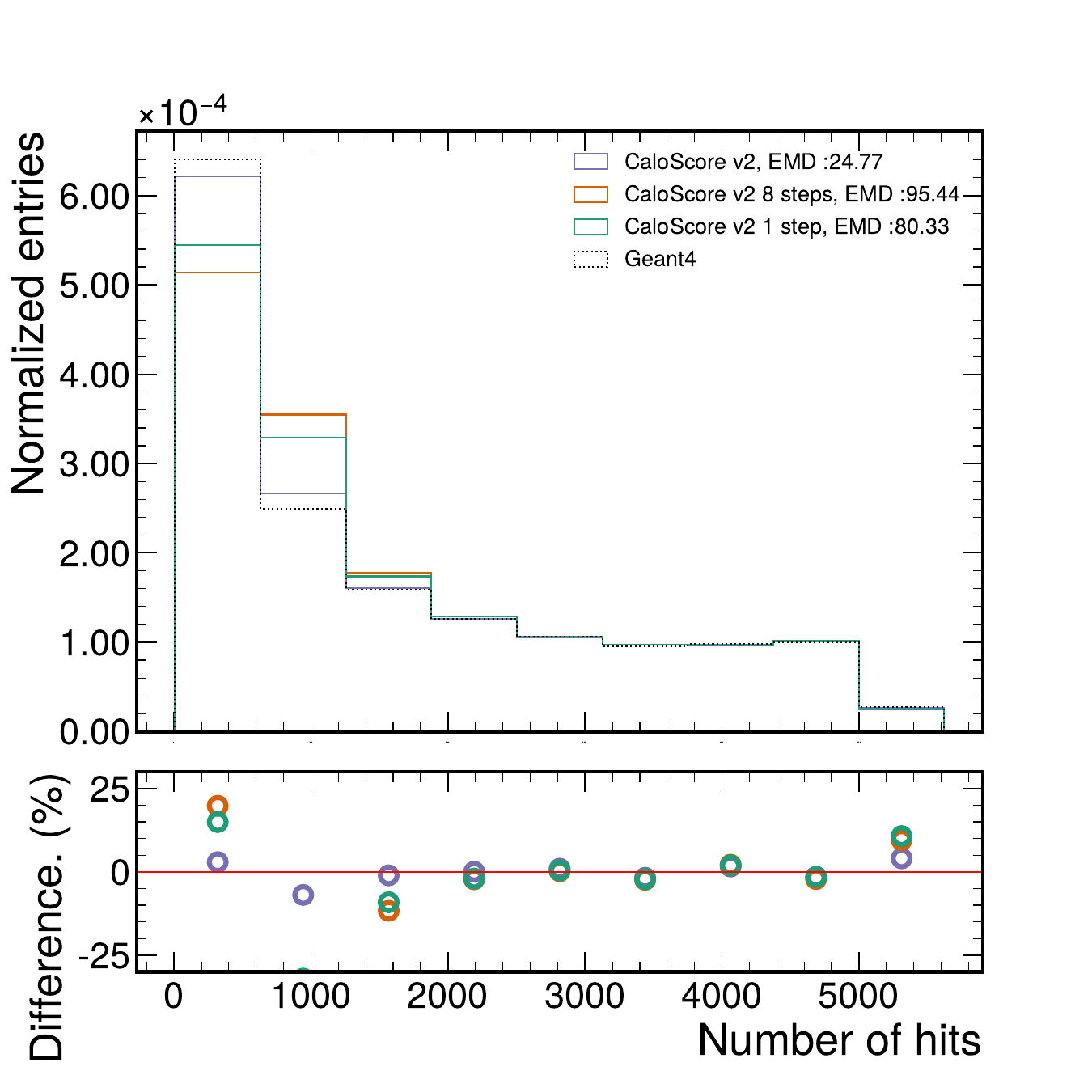}
    \includegraphics[width=0.3\textwidth]{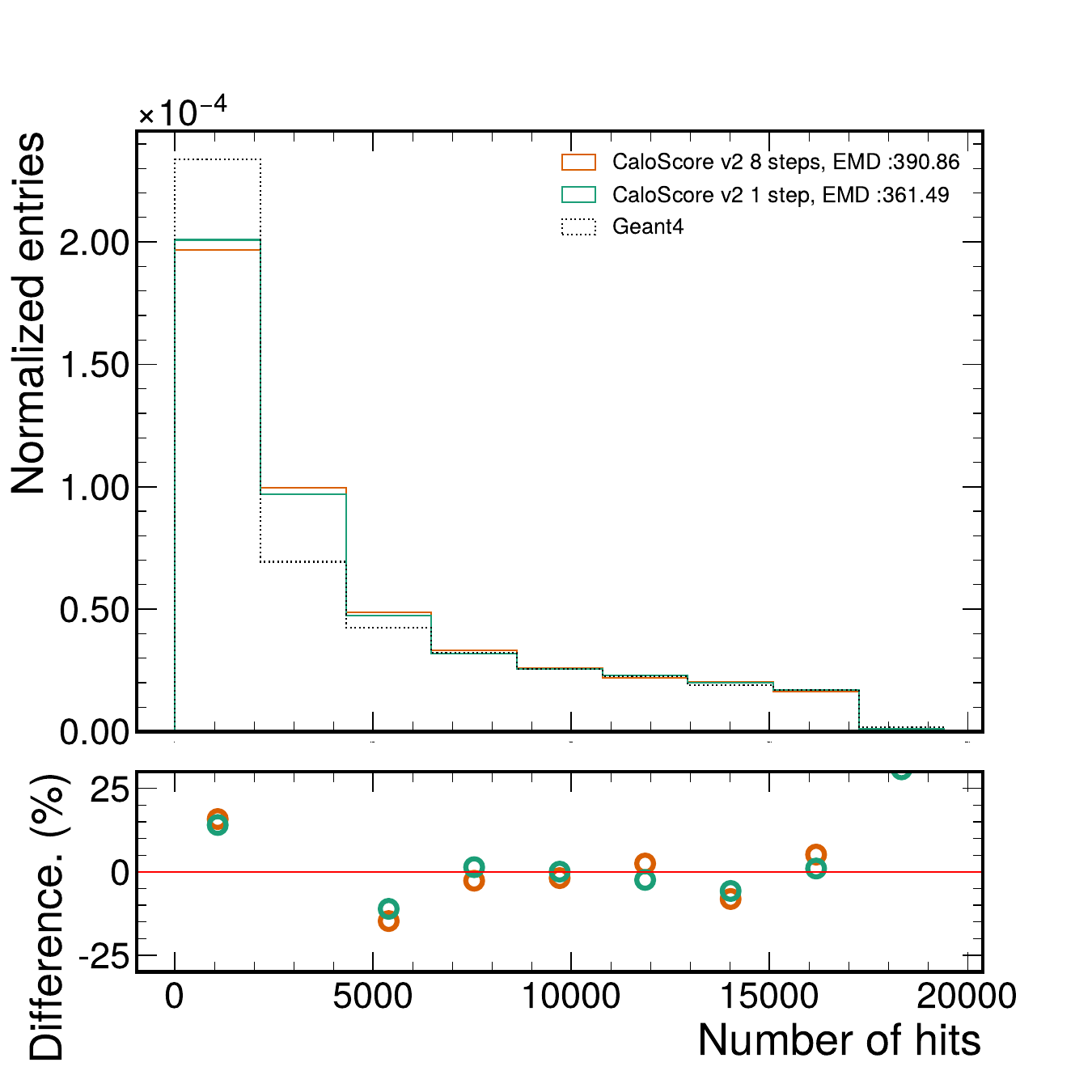}        
\caption{Comparison of the sum of all voxel energies (top) and number of hits (bottom) for datasets 1 (left), 2 (middle), and 3 (right).  The Earth mover's distance (EMD) between each distribution and the \geant~distribution is also provided.}
\label{fig:etot_nhits}
\end{figure*}

Distributions of the total energy deposition and number of calorimeter hits are presented in Fig.~\ref{fig:etot_nhits}. A hit both in the \geant~ simulation and from generated samples is defined by any energy deposition above 0.1 keV in dataset 1 and 15.1 kev in datasets 2 and 3.

We observe a good agreement between generated samples from \calosctwo~ compared to the full simulation. In particular, \calosctwo~after the distillation to a single time step is still able to retain precision. In dataset 1, energies are generated in specific energy intervals, leading to discontinuous values of energy depositions. We compare the results for the total deposited energy with between \calosc~ and \calosctwo~ using the 1-Wasserstein distance, referred as the Earth mover's distance (EMD), between generated samples and the \geant~simulation. Results for the EMD obtained using the total deposited energy are listed in Table.~\ref{tab:EMD}, with Wasserstein GAN (WGAN-GP) and \calosc~values taken from the best performing models presented in \cite{mikuni:caloscore}. 

\begin{table}[ht]
    \centering
	\small
    \caption{Comparison of the earth mover's distance calculated using the total deposited energy. Values for \calosc~are selected from the best performing model reported in \cite{mikuni:caloscore}.}
    \label{tab:EMD}
	\begin{tabular}{l|c|c|c|c|c|cc}
        Model &   \multicolumn{3}{c}{EMD}\\
        &  {\scriptsize dataset 1} & {\scriptsize dataset 2} & {\scriptsize dataset 3} \\
        \hline     
        \calosc &   1.52 & 1.8 & 3.17 \\
        WGAN-GP &   21.55 & 5.95 & 13.29\\
        \calosctwo & 0.21  & 0.13 & - \\ 
        \calosctwo~8 steps & 0.33  & 0.15 & 0.25\\
        \calosctwo~1 step& 0.35 & 0.19 & 0.40\\       
	\end{tabular}
\end{table}

We observe a significant improvement compared to \calosc~ with EMD values decreasing by a factor 10 compared to previous results. This improvement stems from the independent determination of the energy depositions per layer achieved by the separate diffusion process. Reduced fidelity is observed in the distilled models compared to baseline \calosctwo. Nevertheless, even the single-shot model still improves upon \calosc.

\begin{figure*}[ht]
\centering
    \includegraphics[width=0.3\textwidth]{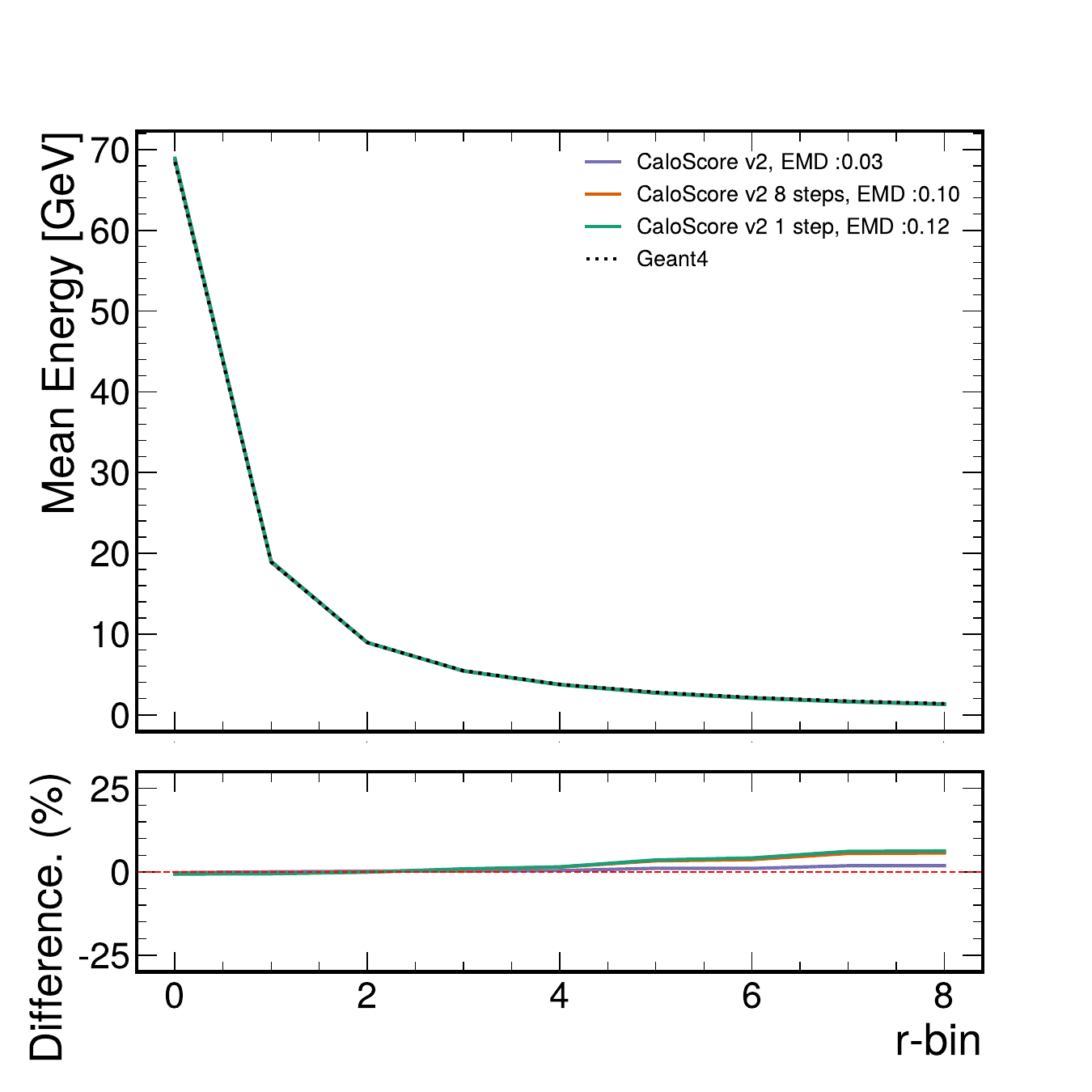}
    \includegraphics[width=0.3\textwidth]{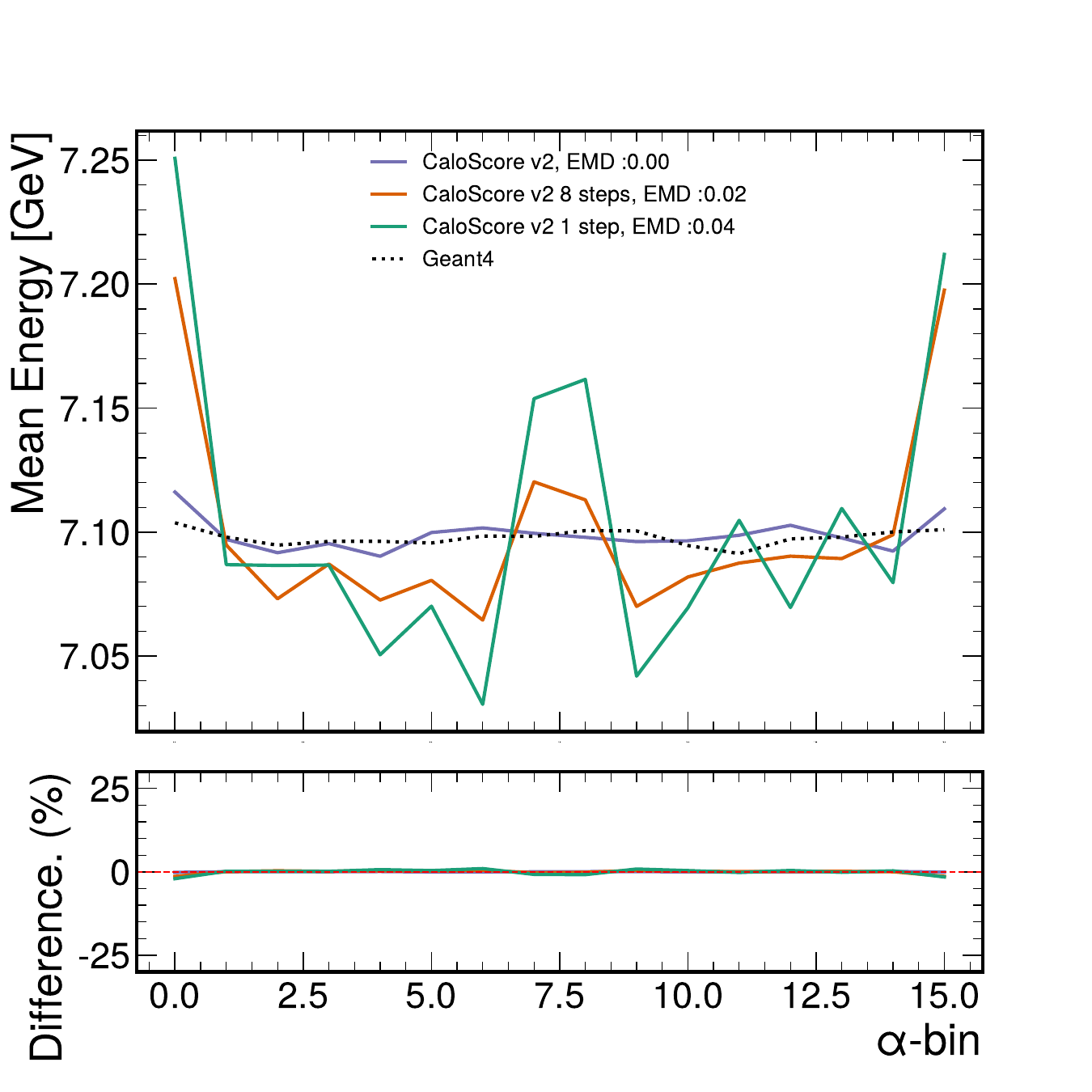}
    \includegraphics[width=0.3\textwidth]{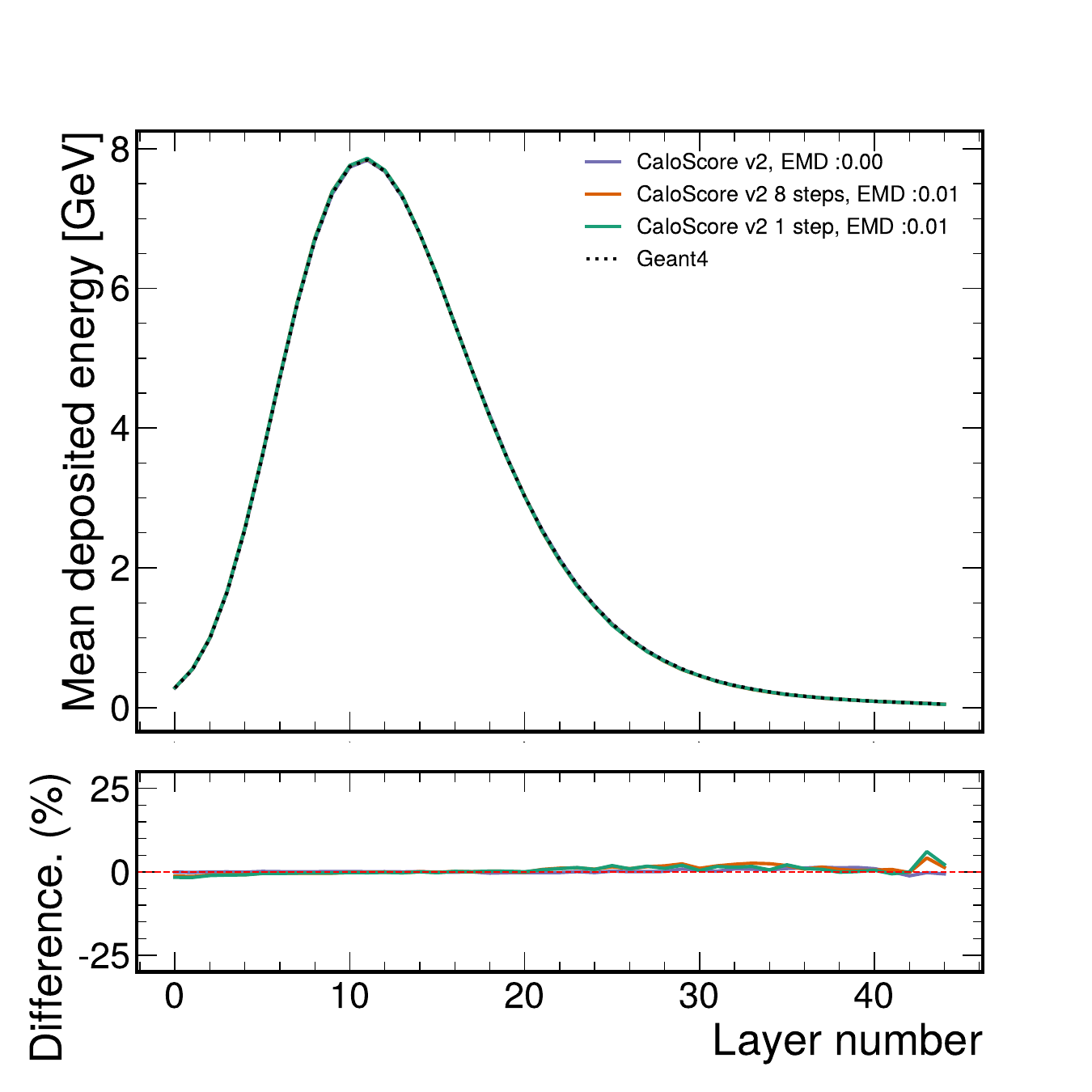}
    \includegraphics[width=0.3\textwidth]{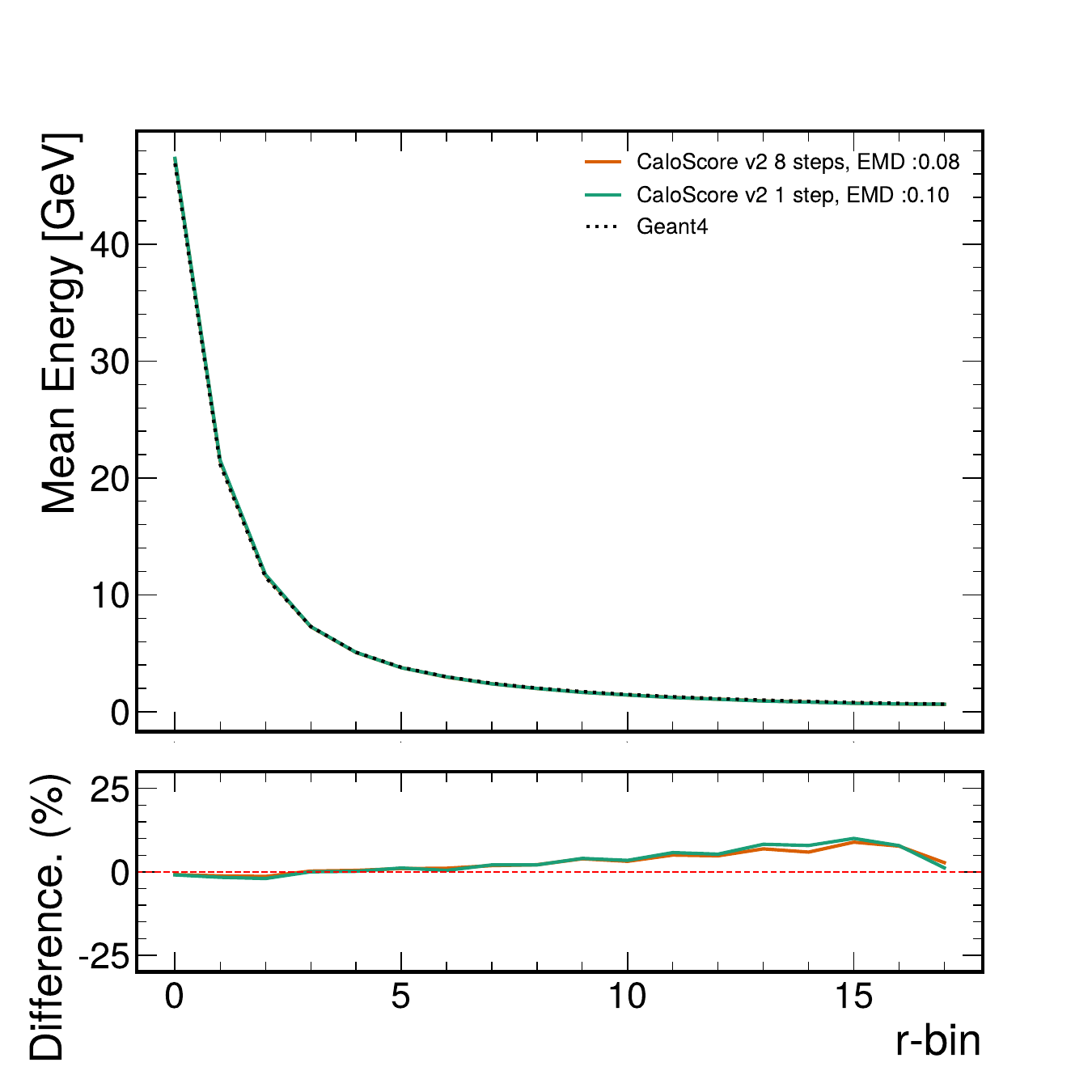}
    \includegraphics[width=0.3\textwidth]{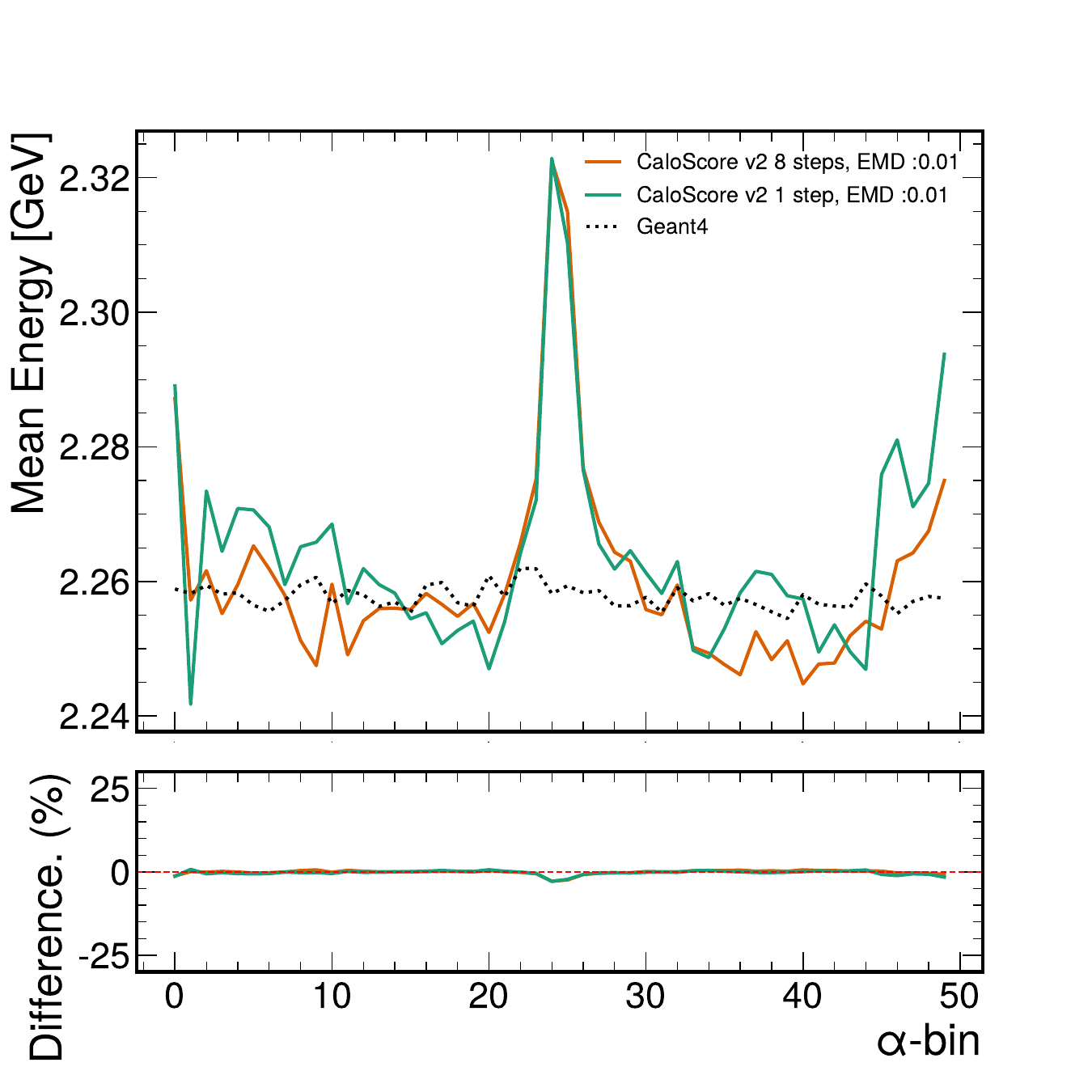}
    \includegraphics[width=0.3\textwidth]{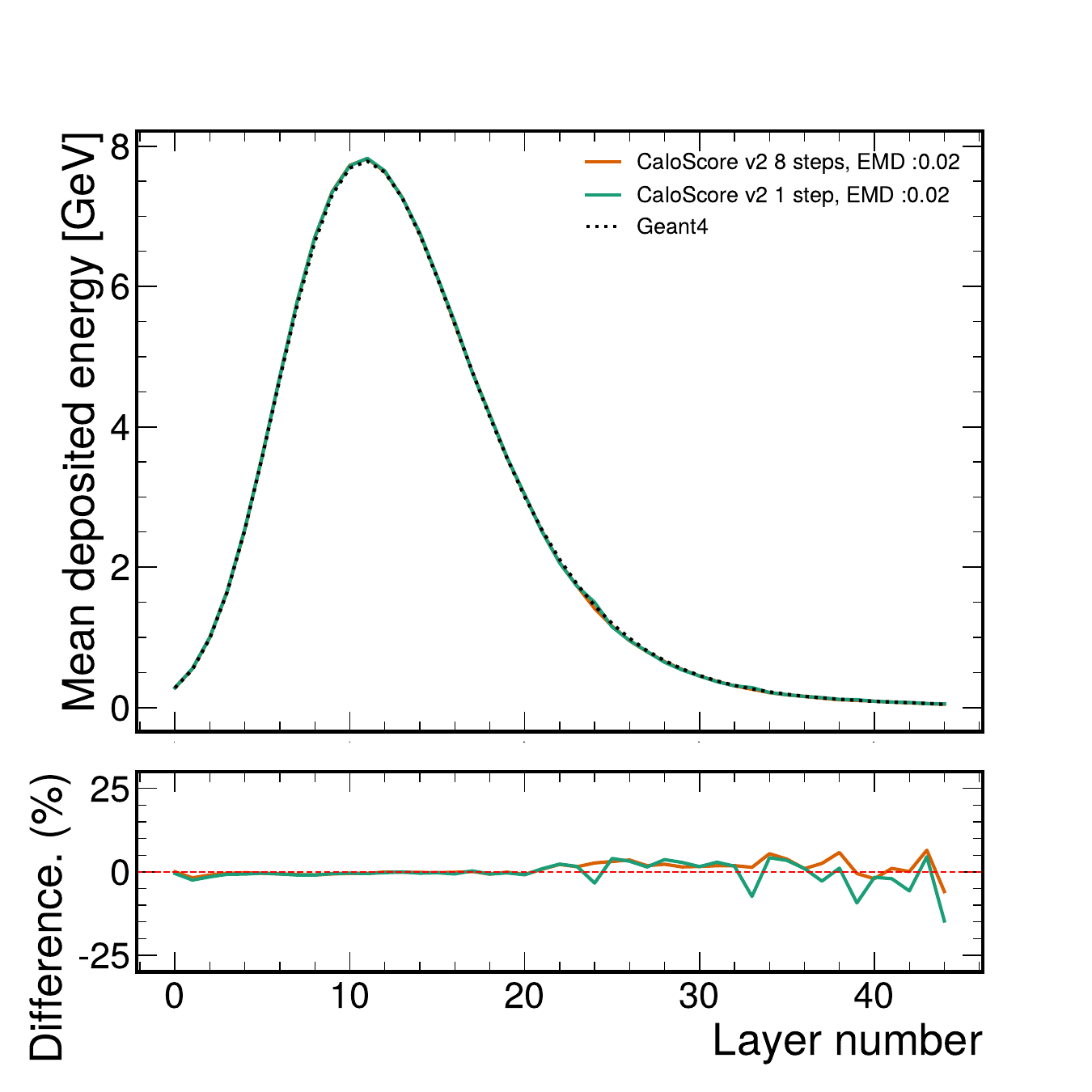}
    
\caption{Comparison of the average deposited energies in the $r$- (left), $\alpha$- (middle), and z-coordinates (right) for datasets 2 (top) and 3 (bottom). The Earth mover's distance (EMD) between each distribution and the \geant~distribution is also provided.}
\label{fig:emean}
\end{figure*}

Next, we study the mean deposited energy versus $r$, $\alpha$, and layer number presented in Fig.~\ref{fig:emean}.

The mean energy as a function of layer number is determined by the independent diffusion model, making it insensitive to the modeling of individual voxels. In contrast, the distributions concerning $r$ and $\alpha$ are sensitive to the modeling of the individual voxels, with agreement within 10\% observed in all distillation levels.

\begin{figure*}[ht]
\centering
    \includegraphics[width=0.22\textwidth]{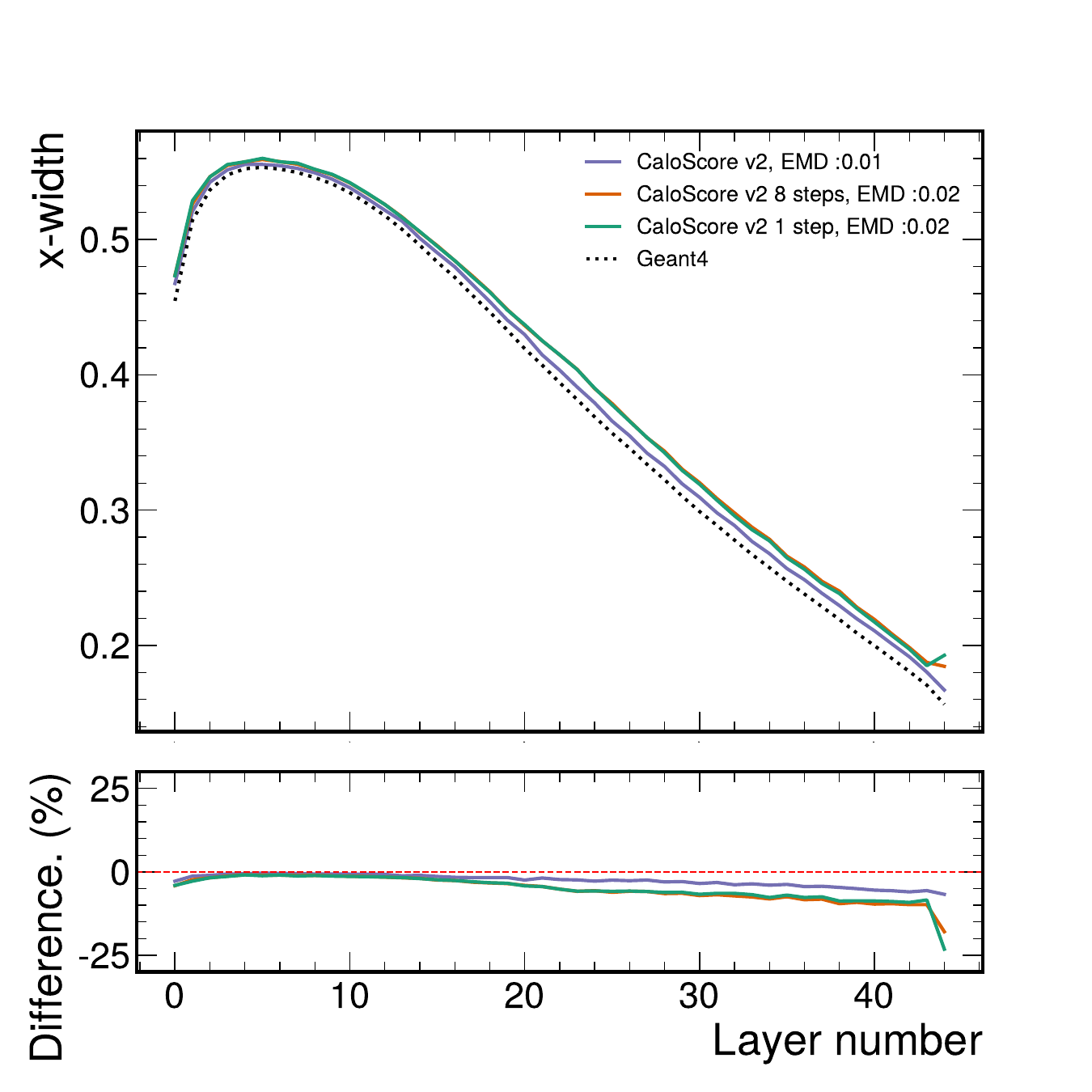}
    \includegraphics[width=0.22\textwidth]{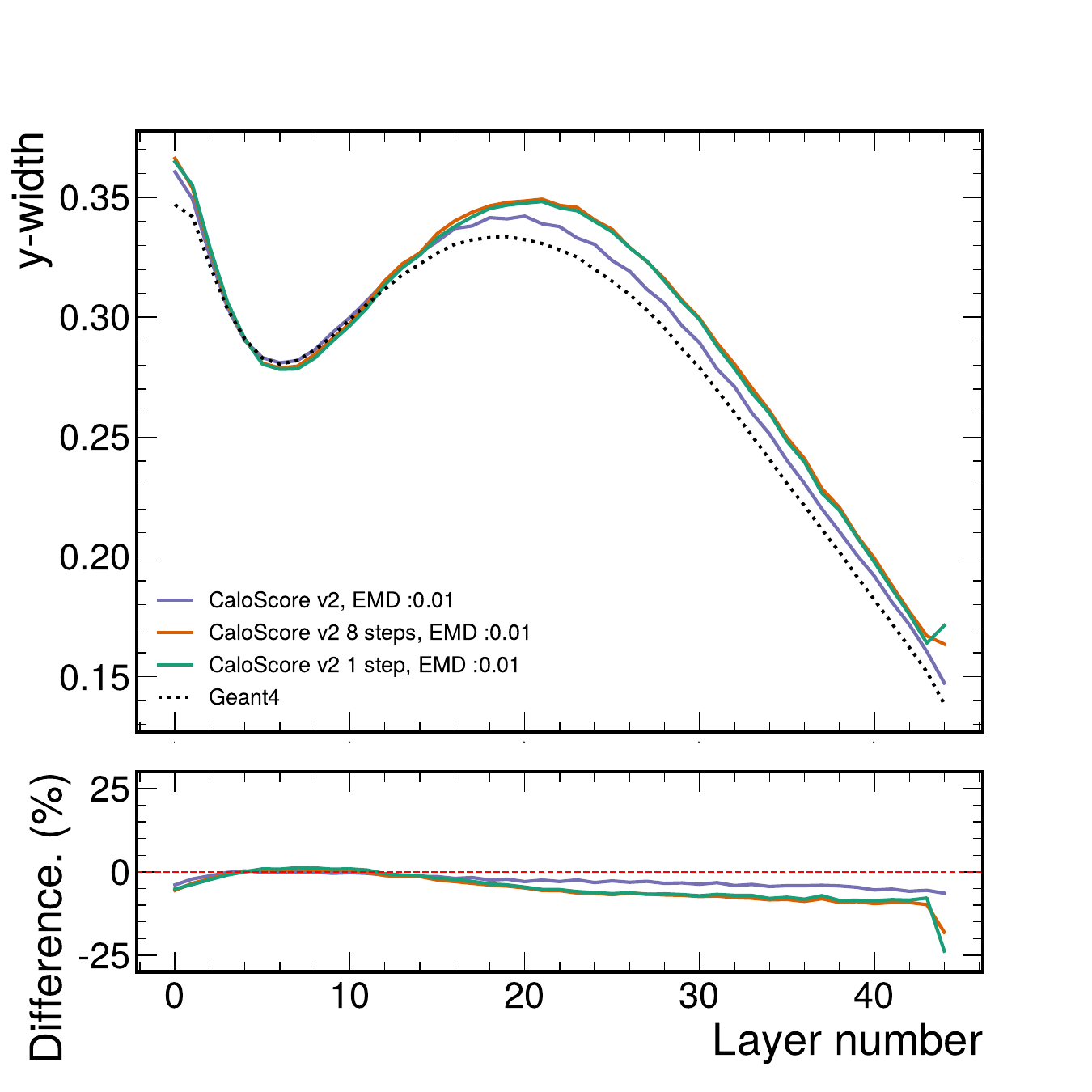}
    \includegraphics[width=0.22\textwidth]{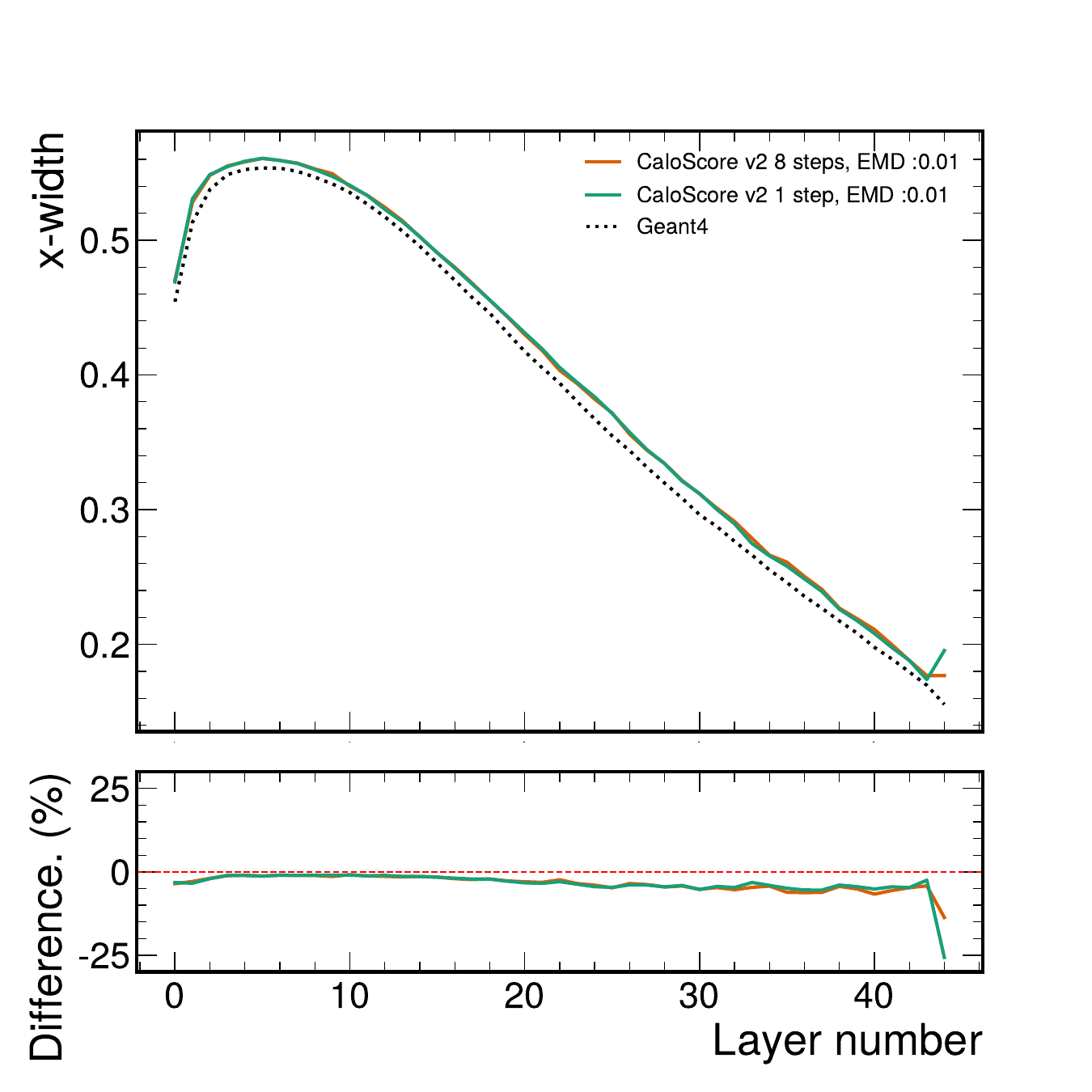}
    \includegraphics[width=0.22\textwidth]{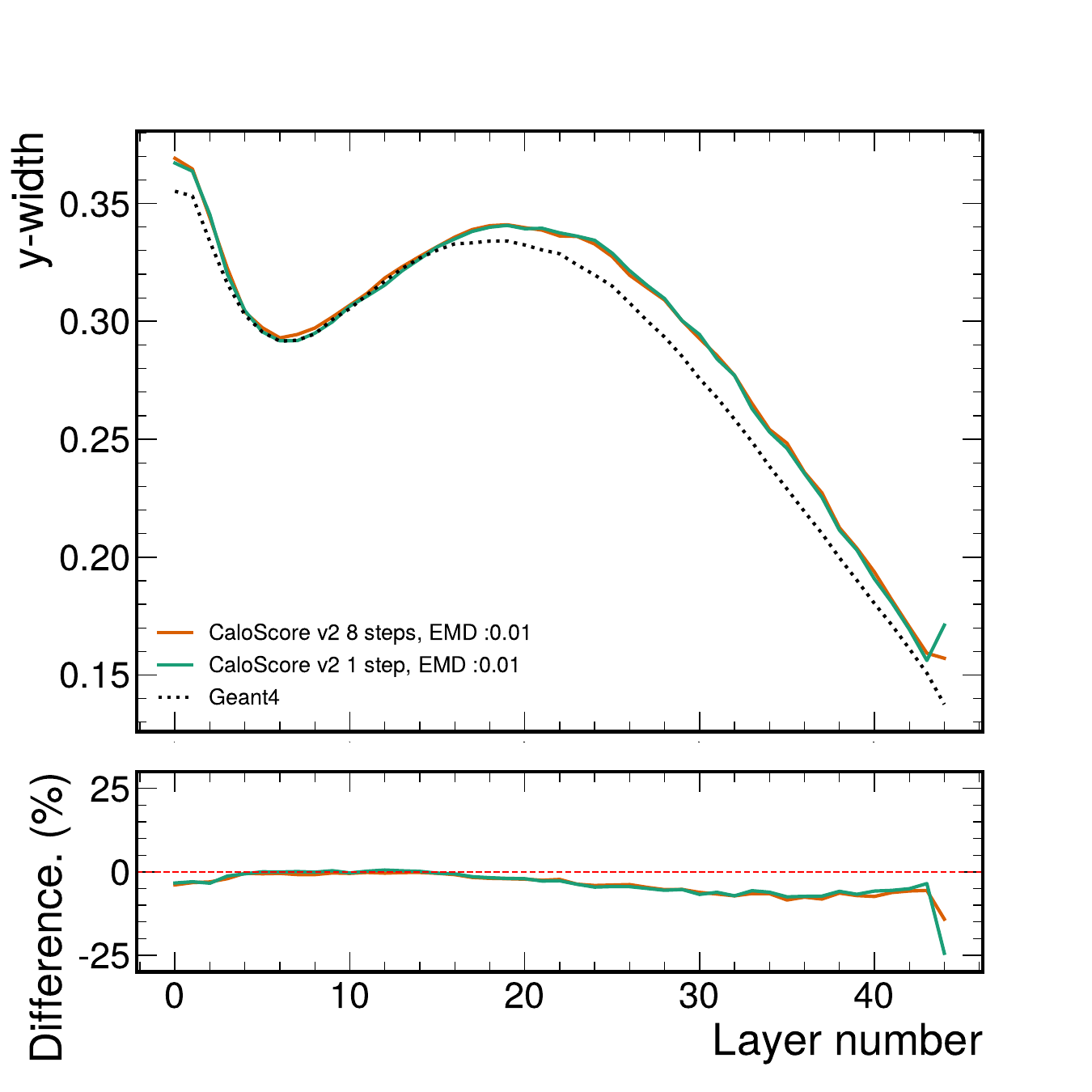}
    
\caption{Comparison of the particle shower width in the $r$-  and $\alpha$-  directions in datasets 2 (first two figures from the left) and 3 (last two figures from the left).  The Earth mover's distance (EMD) between each distribution and the \geant~distribution is also provided.}
\label{fig:shower_w}
\end{figure*}

Additionally, we investigate the angular distributions of the calorimeter showers in datasets 2 and 3 in terms of the shower width, shown in Fig.~\ref{fig:shower_w}.  The shower width $\sigma_i$ with $x_i, i\in [1,2]$ representing the $r$- and $\alpha$- coordinates is calculated as: 
\begin{equation}
    \sigma_i = \sqrt{\left< x_i^2 \right > - \left< x_i \right >^2 },
\end{equation}
with energy-weighted mean defined as
\begin{equation}
    \left< x_i \right > =  \frac{\sum_j x_{i,j} E_j}{\sum_j E_j}.
\end{equation}

The agreement of \calosctwo~is often within 10\% compared to the \geant~ simulation, with exception to the tails of the distribution located in the later layers of the detector with differences more pronounced for the distilled models compared to the baseline \calosctwo.

\begin{figure*}[ht]
\centering
    \includegraphics[width=0.23\textwidth]{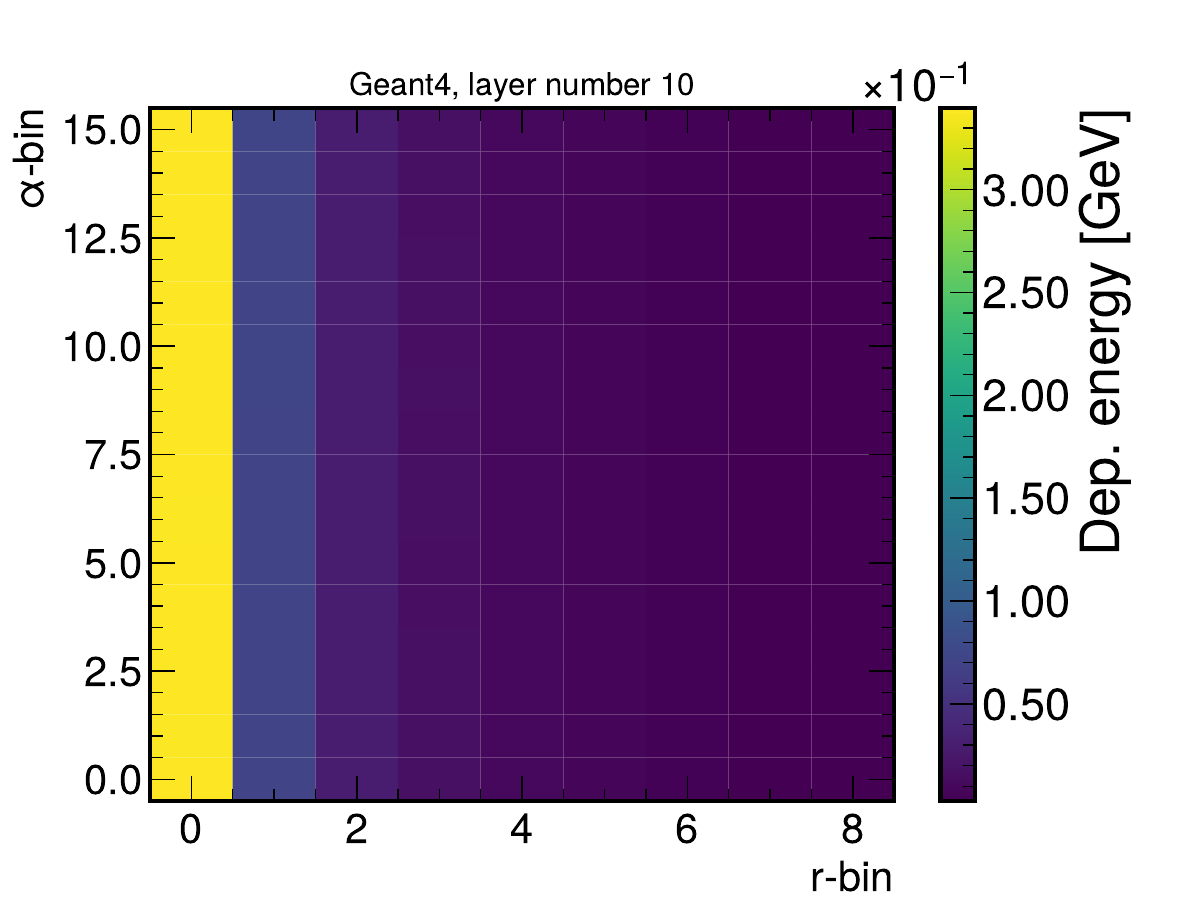}
    \includegraphics[width=0.23\textwidth]{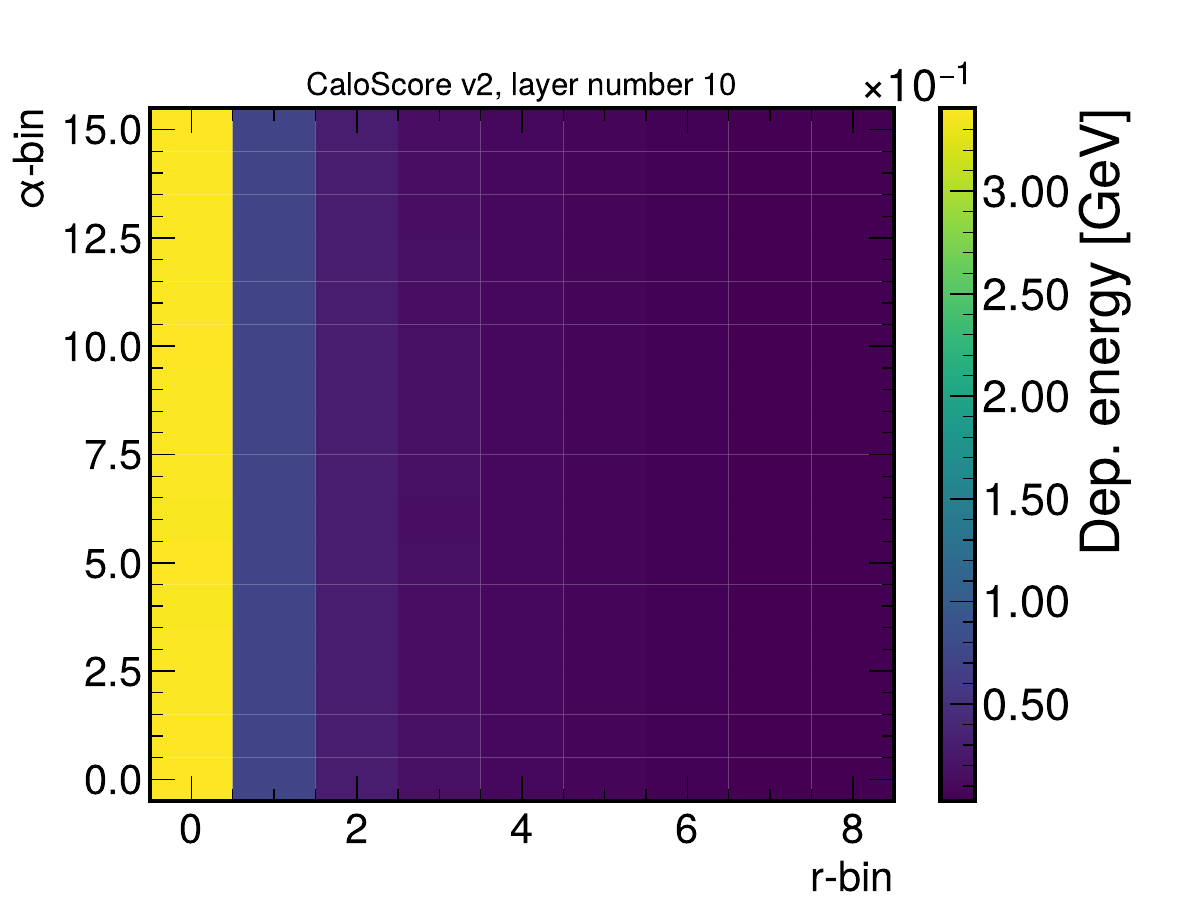}
    \includegraphics[width=0.23\textwidth]{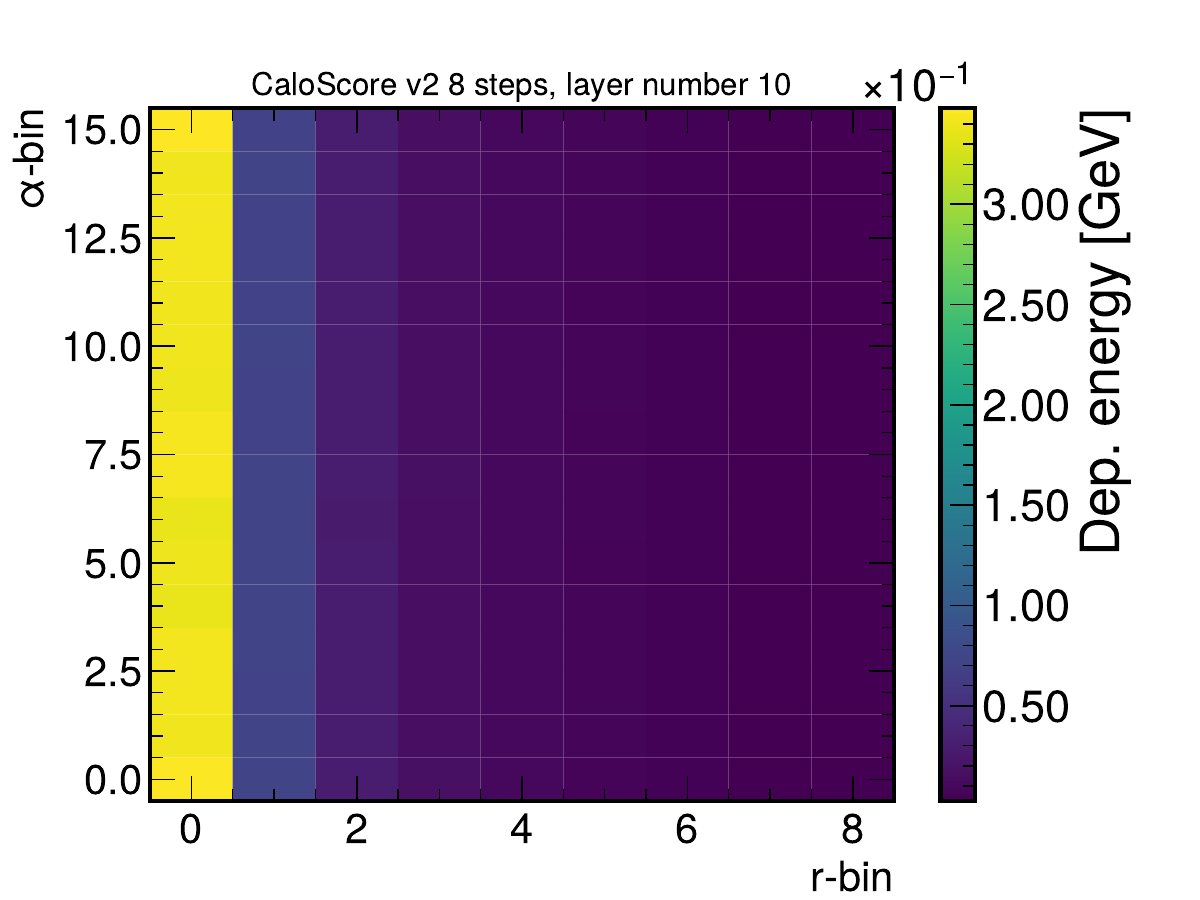}
    \includegraphics[width=0.23\textwidth]{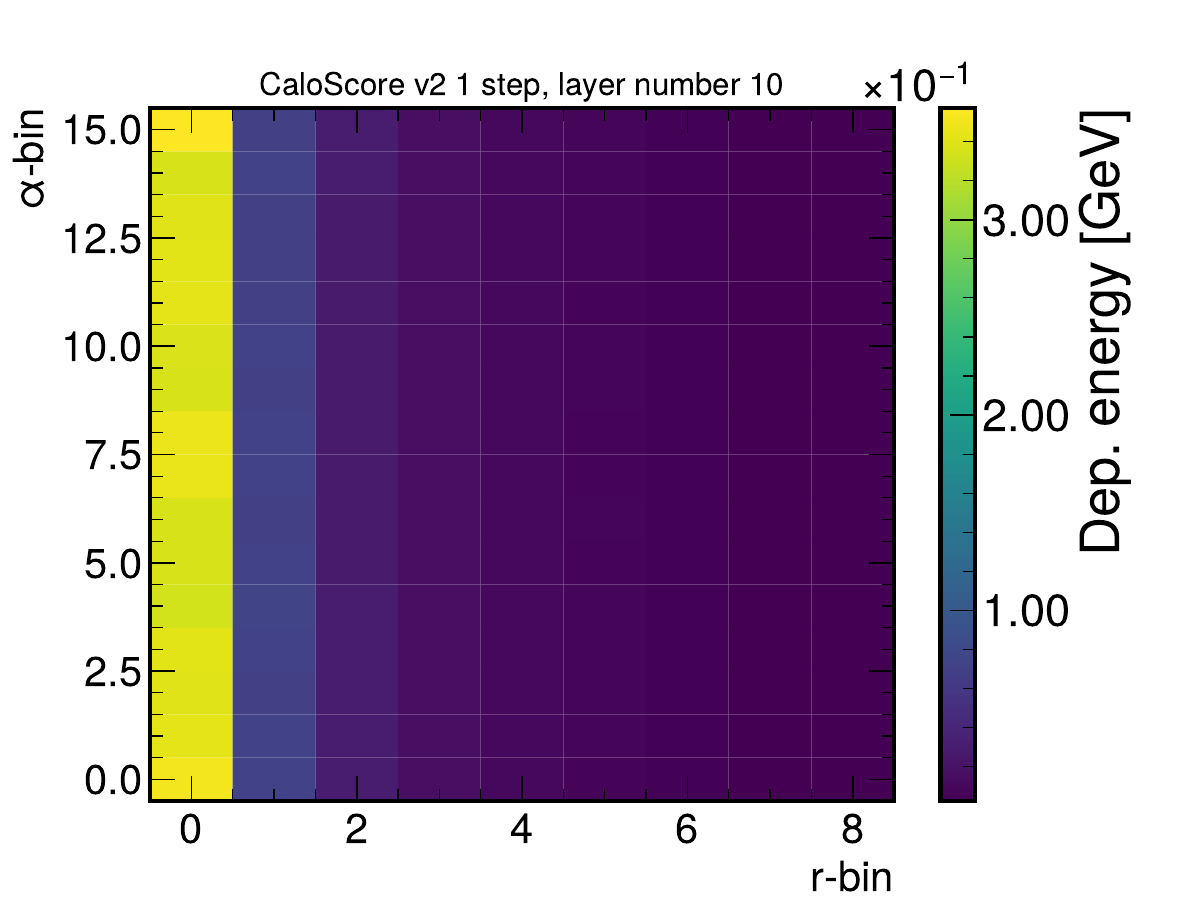}
    \includegraphics[width=0.23\textwidth]{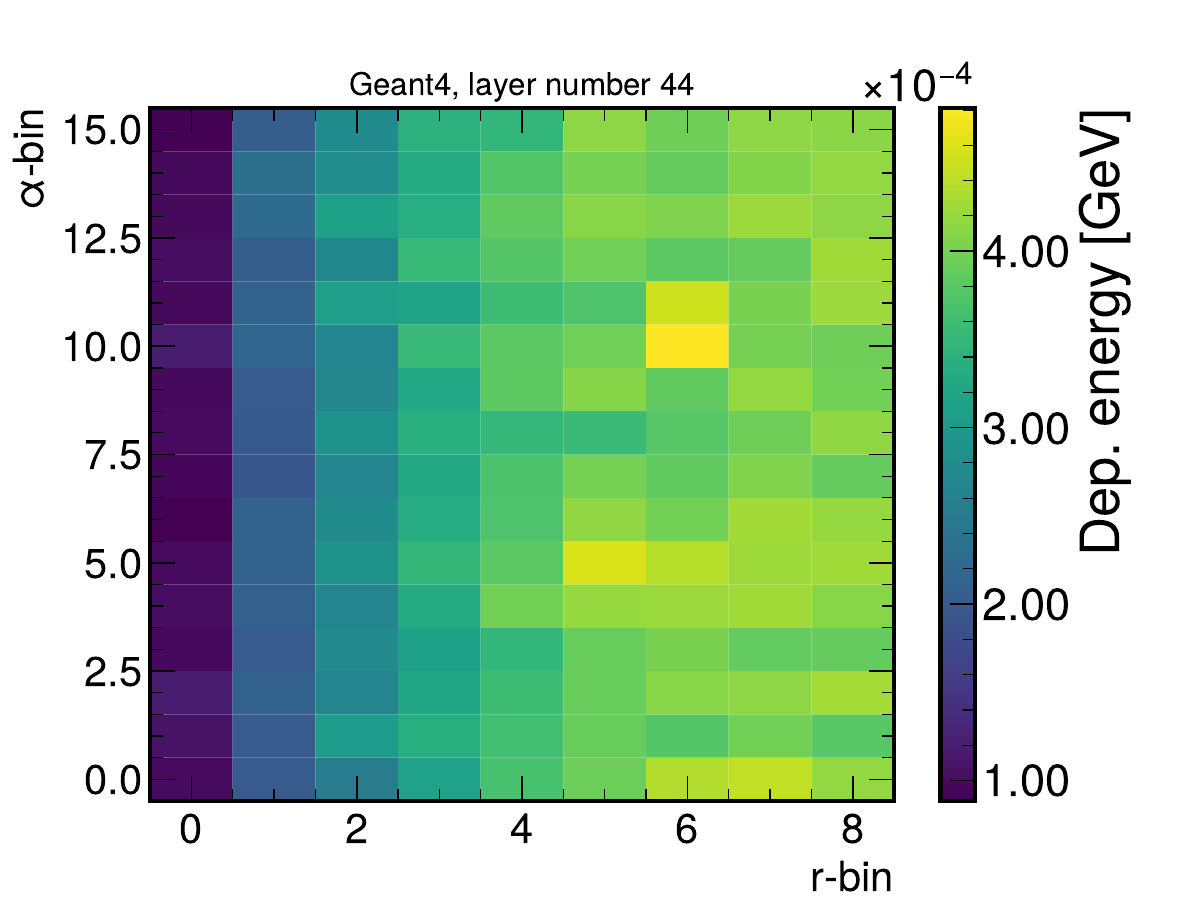}
    \includegraphics[width=0.23\textwidth]{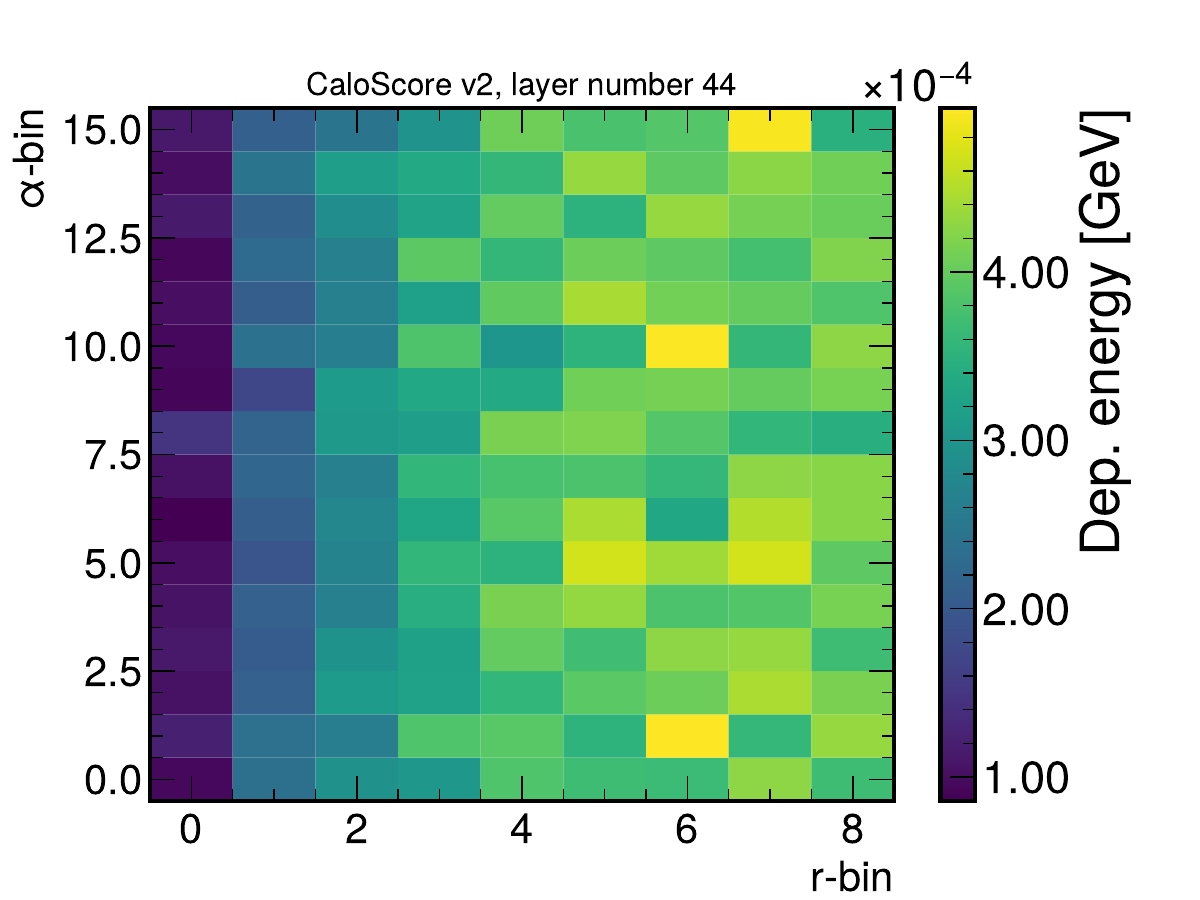}
    \includegraphics[width=0.23\textwidth]{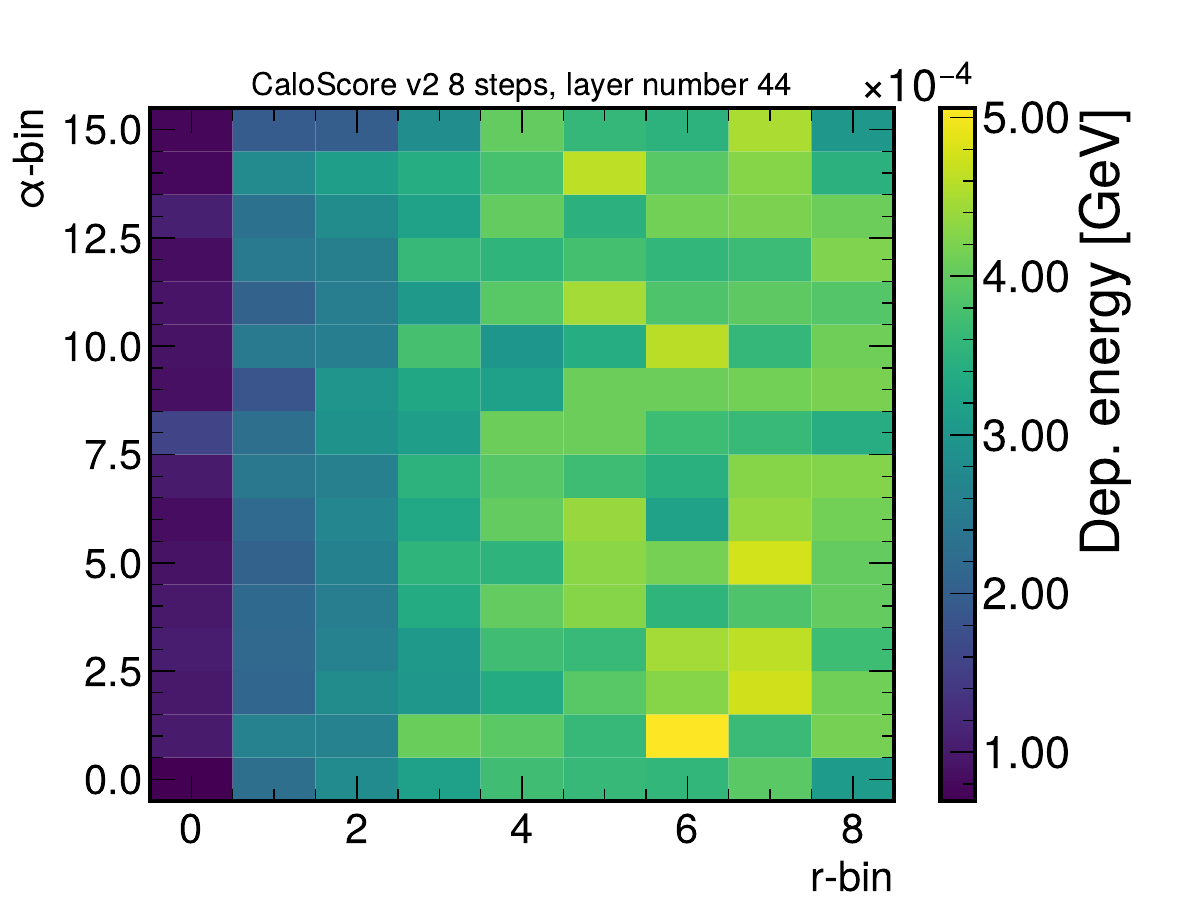}
    \includegraphics[width=0.23\textwidth]{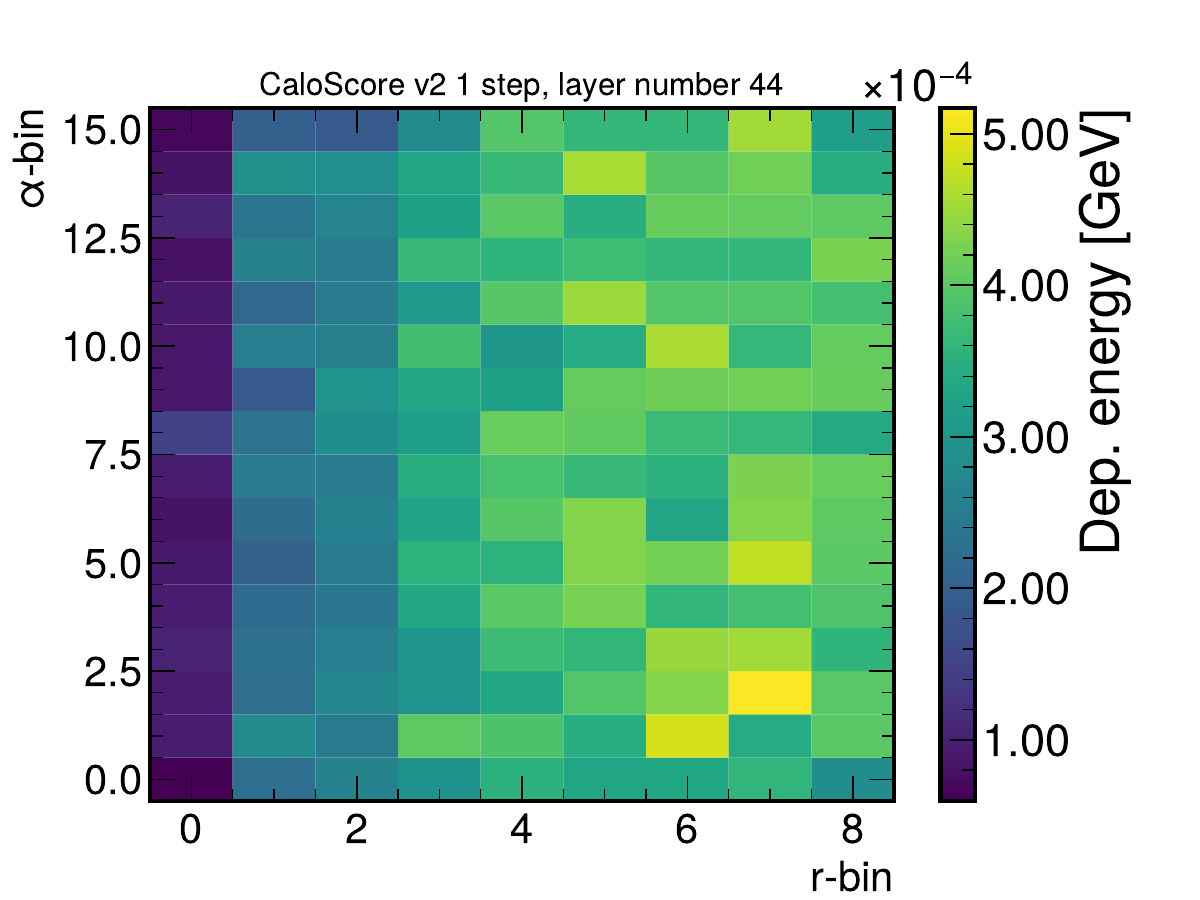}
    \includegraphics[width=0.32\textwidth]{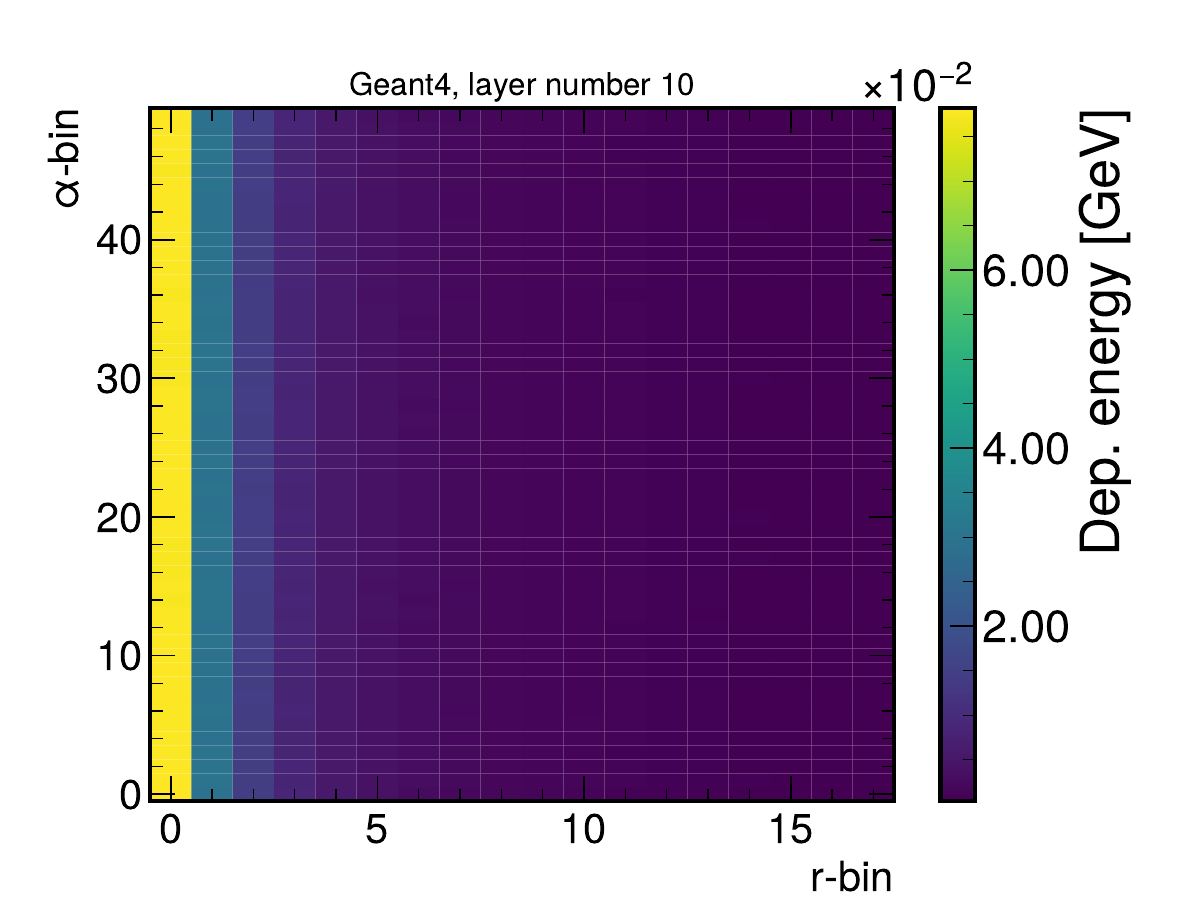}
    \includegraphics[width=0.32\textwidth]{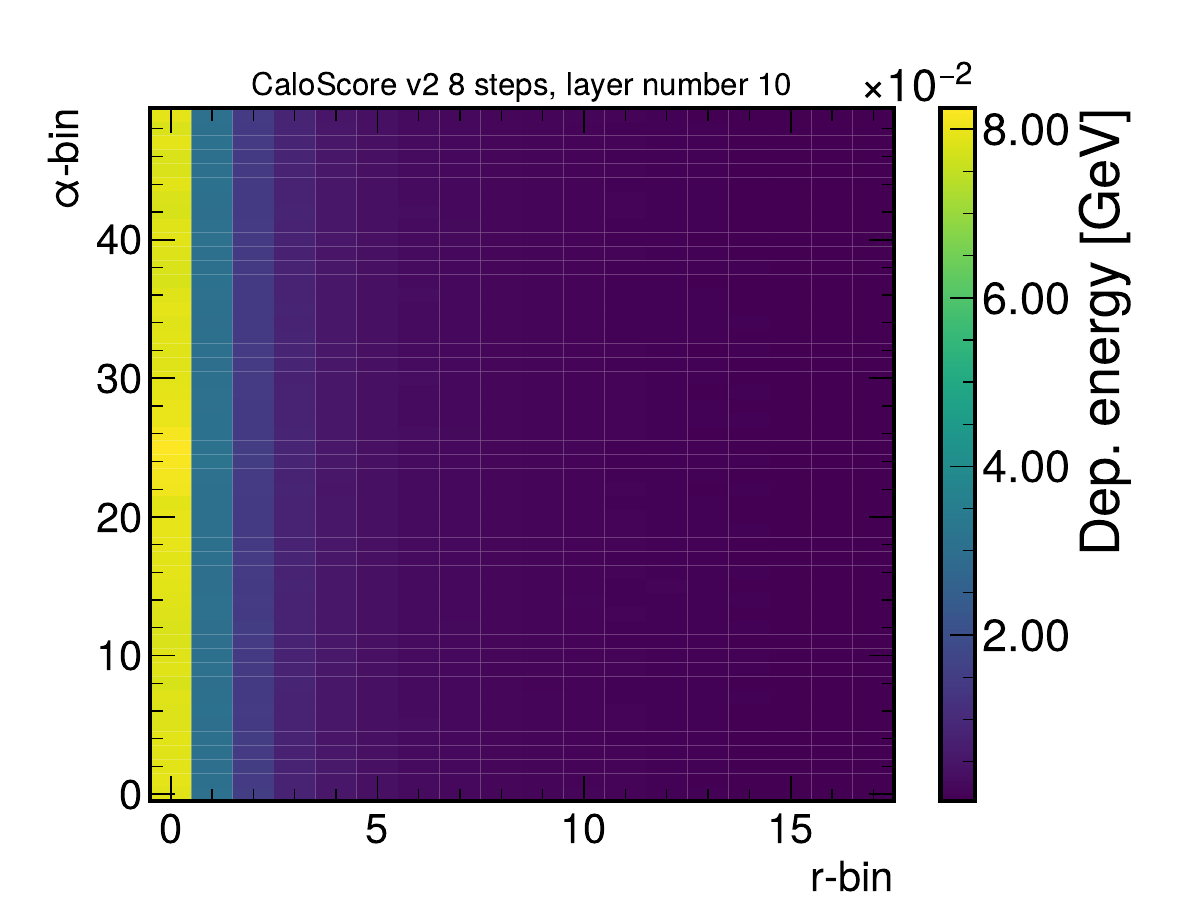}
    \includegraphics[width=0.32\textwidth]{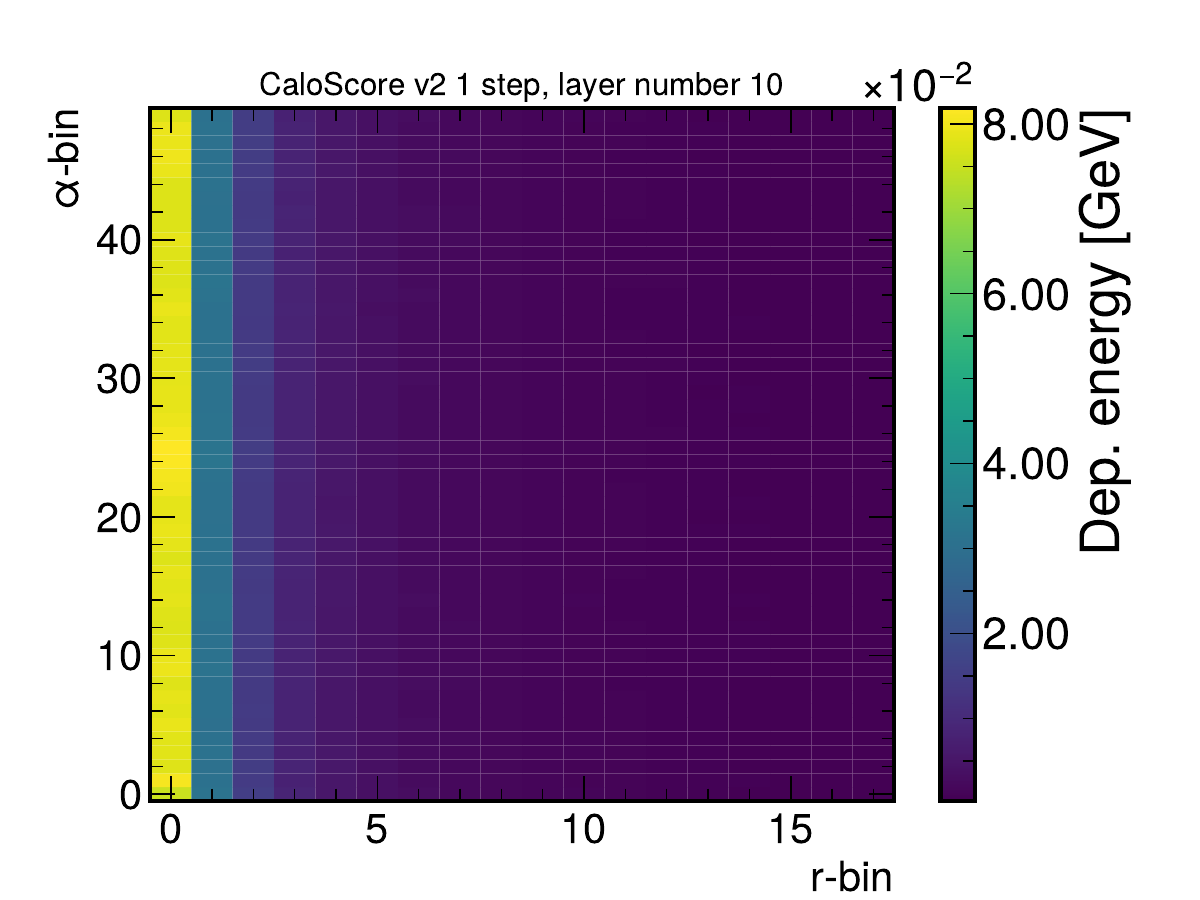}
    \includegraphics[width=0.32\textwidth]{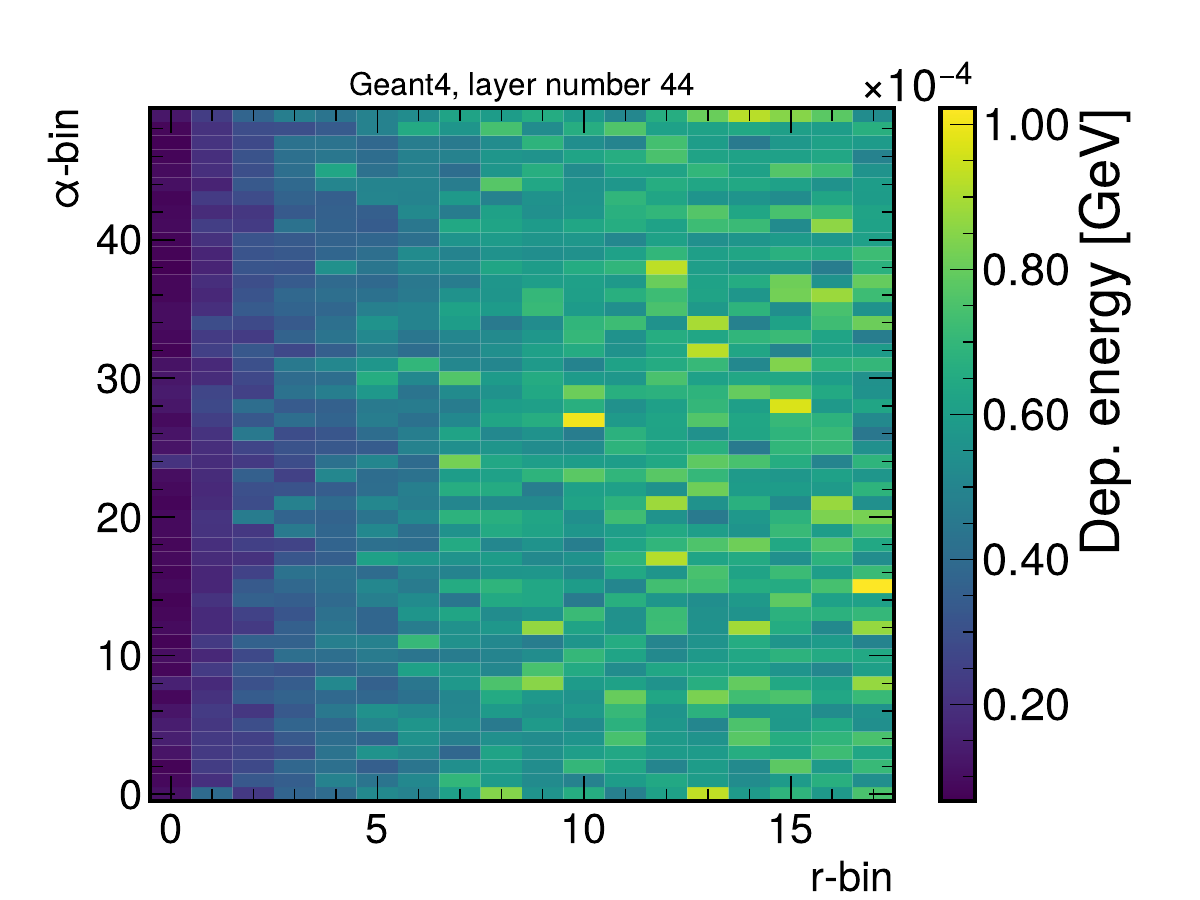}
    \includegraphics[width=0.32\textwidth]{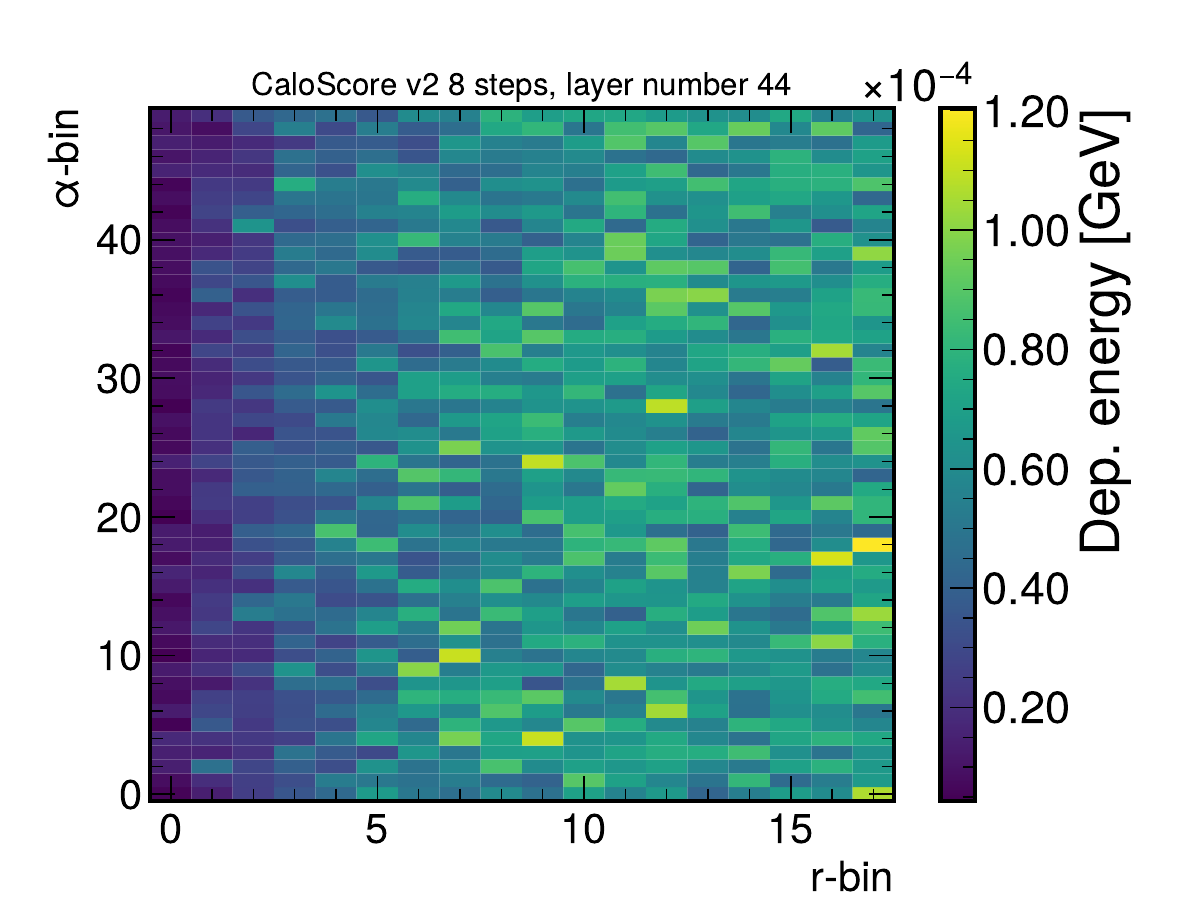}
    \includegraphics[width=0.32\textwidth]{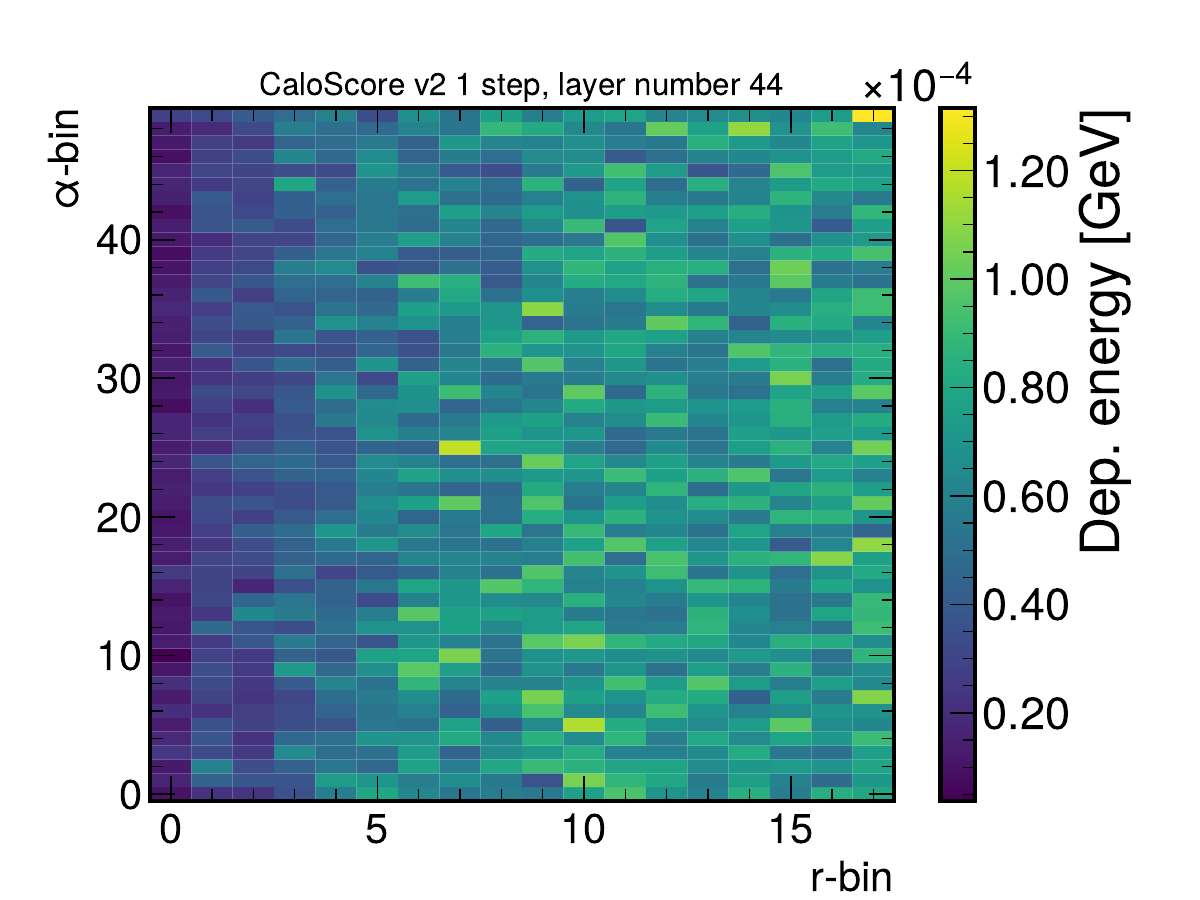}

\caption{The 2-dimensional distribution of the mean deposited energy in layers with highest (first and third rows) and lowest (second and fourth rows) mean energy depositions in datasets 2 (first two rows) and 3 (last two rows). Simulated samples from \geant~are shown in the first column, compared with \calosctwo~using different number of sampling time steps.}
\label{fig:2D_showers}
\end{figure*}

We also perform a visual inspection for datasets 2 and 3 using samples generated by \calosctwo~ by looking at the average energy deposition per voxel for 10,000 showers in layers 10 and 44, the layers with the highest and lowest average energy deposition, respectively. Results are shown in Fig.~\ref{fig:2D_showers}.

In layer 10, the majority of the energy deposition is located near $r=0$ since incident particles are generated at the center and orthogonal to the detector plane.  As the electromagnetic shower evolves, the interactions with the detector material result in more energy  deposited away from the center, with layer 44 showing the majority of the energy depositions spread over higher values of $r$. In all cases, the \calosctwo~ samples are able to reproduce the correct trend and do not seem to create any noticeable mismodeling.

\begin{figure*}[ht]
\centering
    \includegraphics[width=0.31\textwidth]{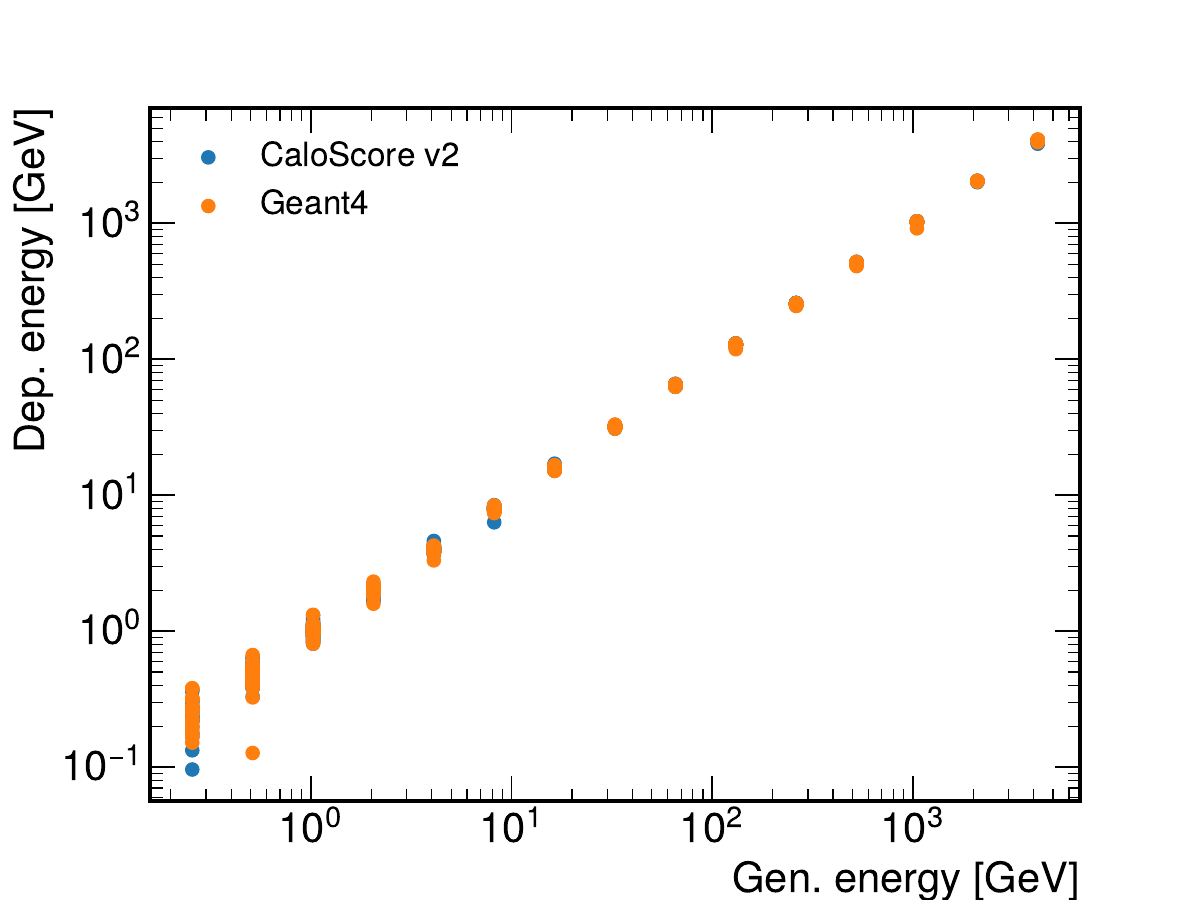}
    \includegraphics[width=0.31\textwidth]{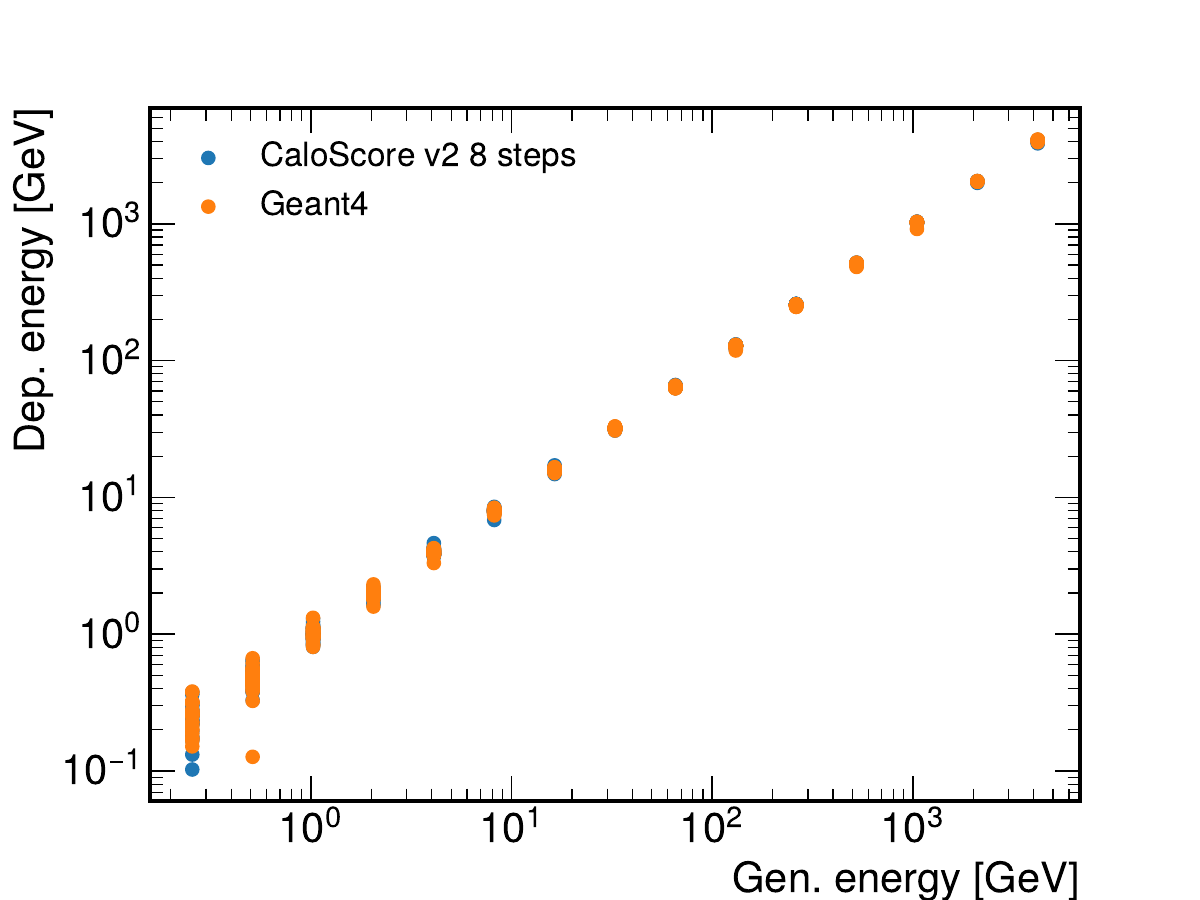}
    \includegraphics[width=0.31\textwidth]{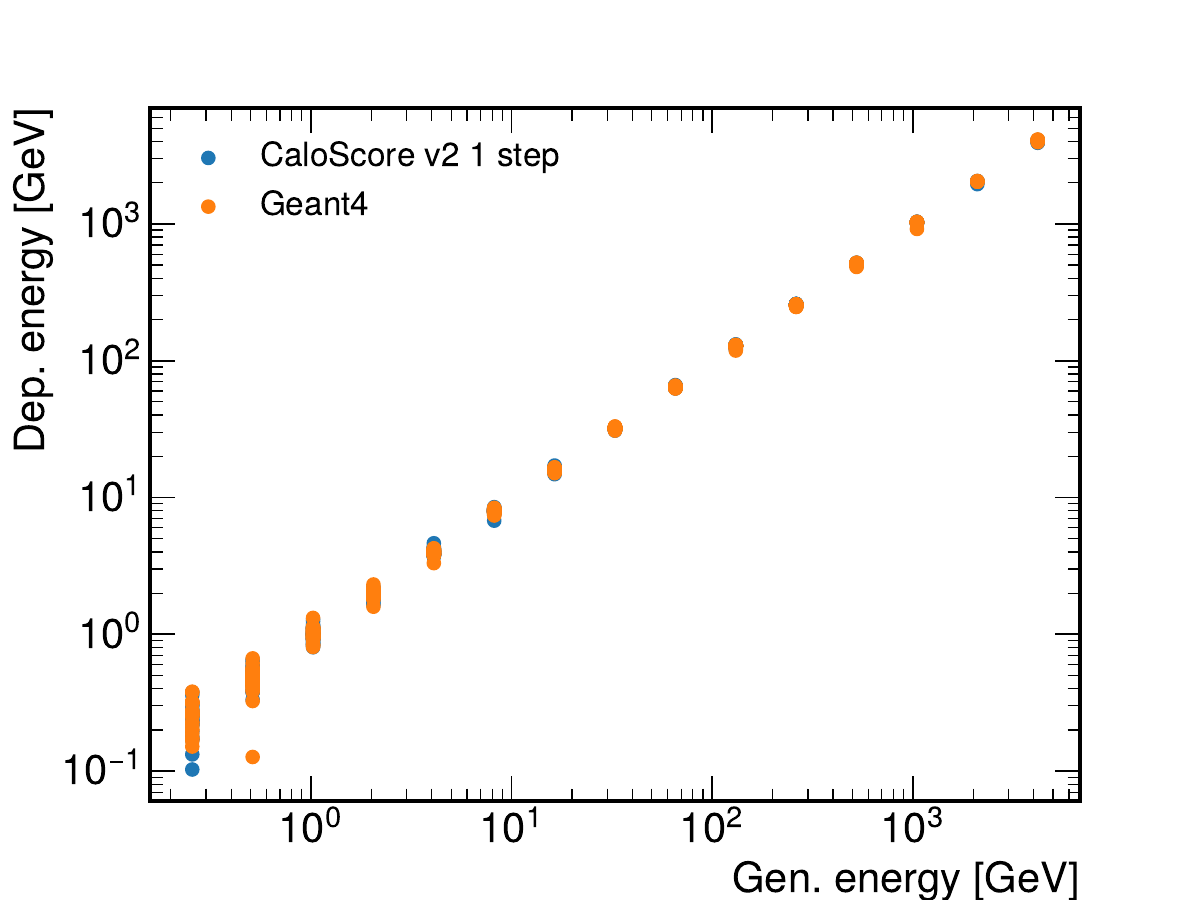}
    \includegraphics[width=0.31\textwidth]{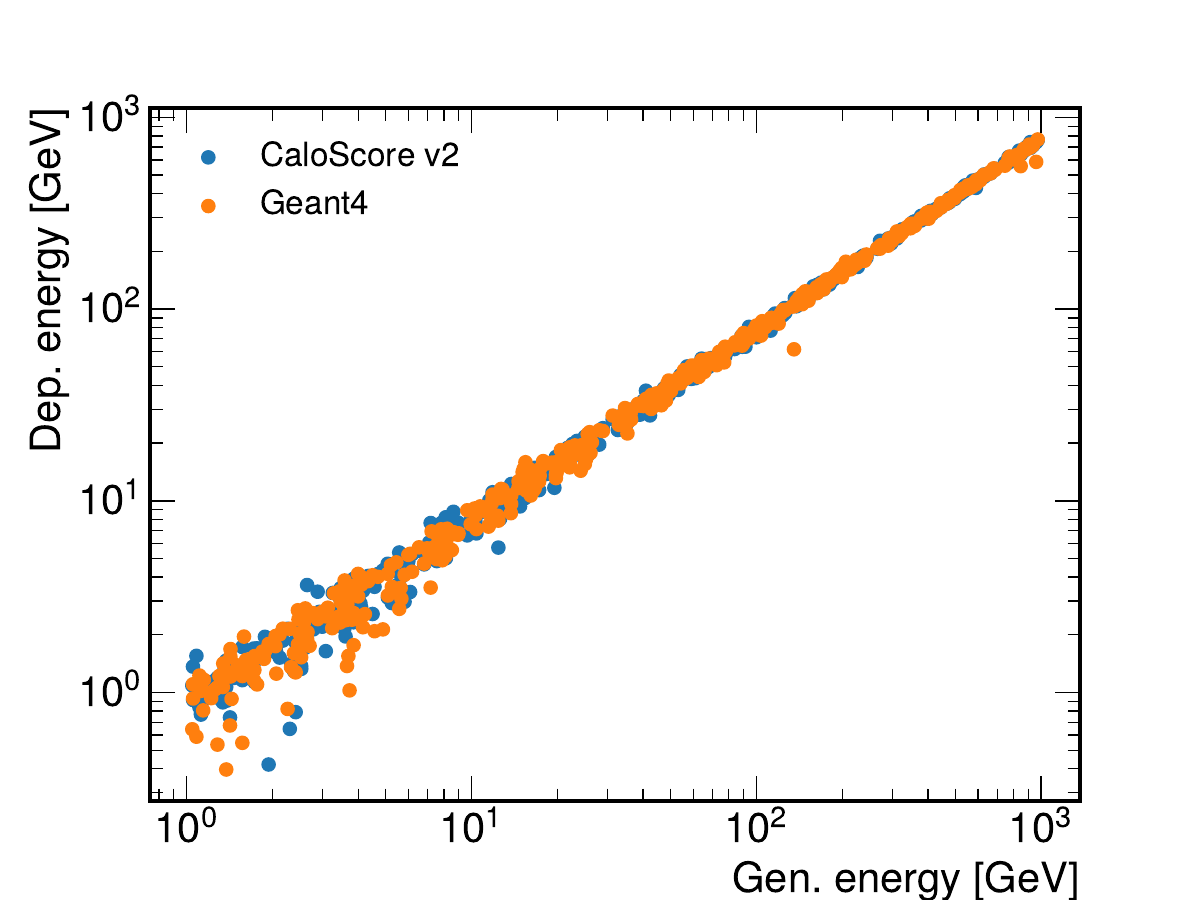}
    \includegraphics[width=0.31\textwidth]{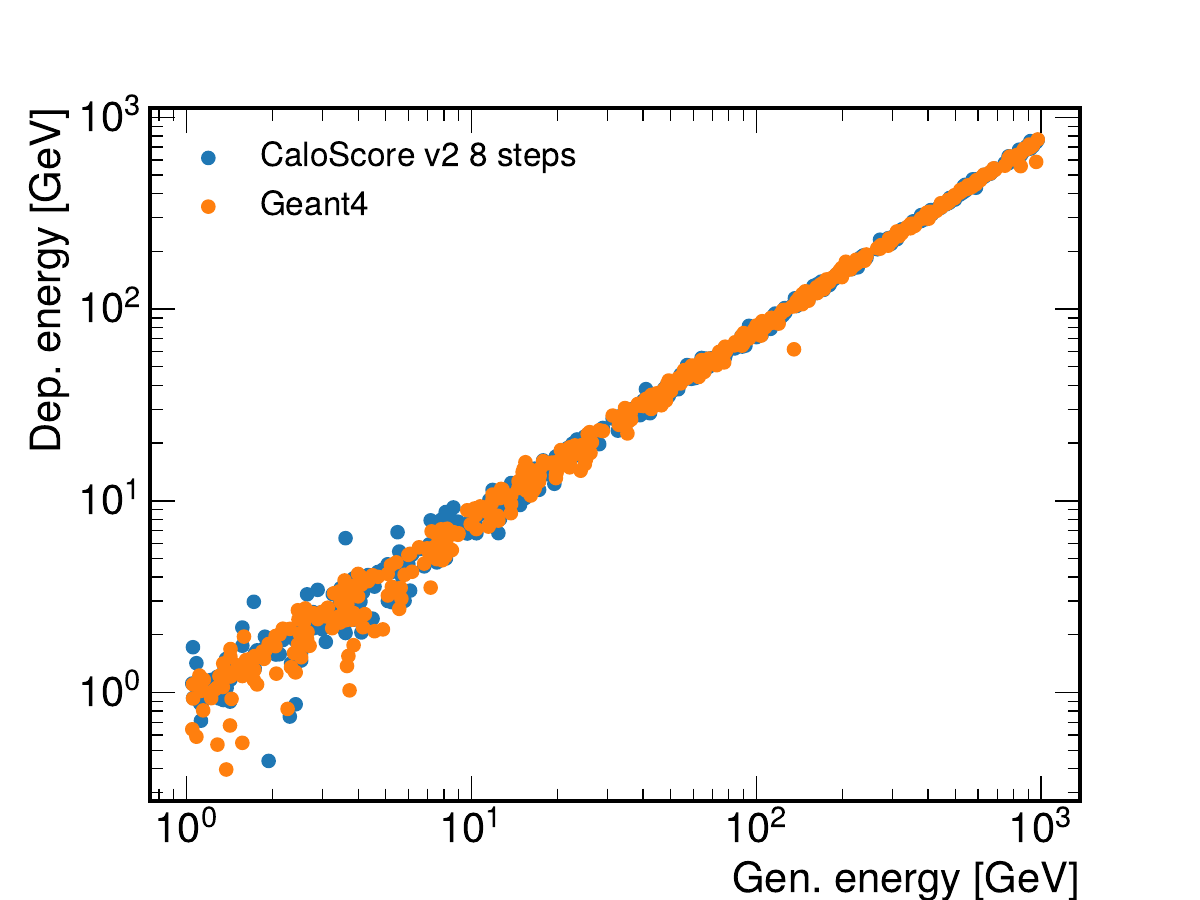}
    \includegraphics[width=0.31\textwidth]{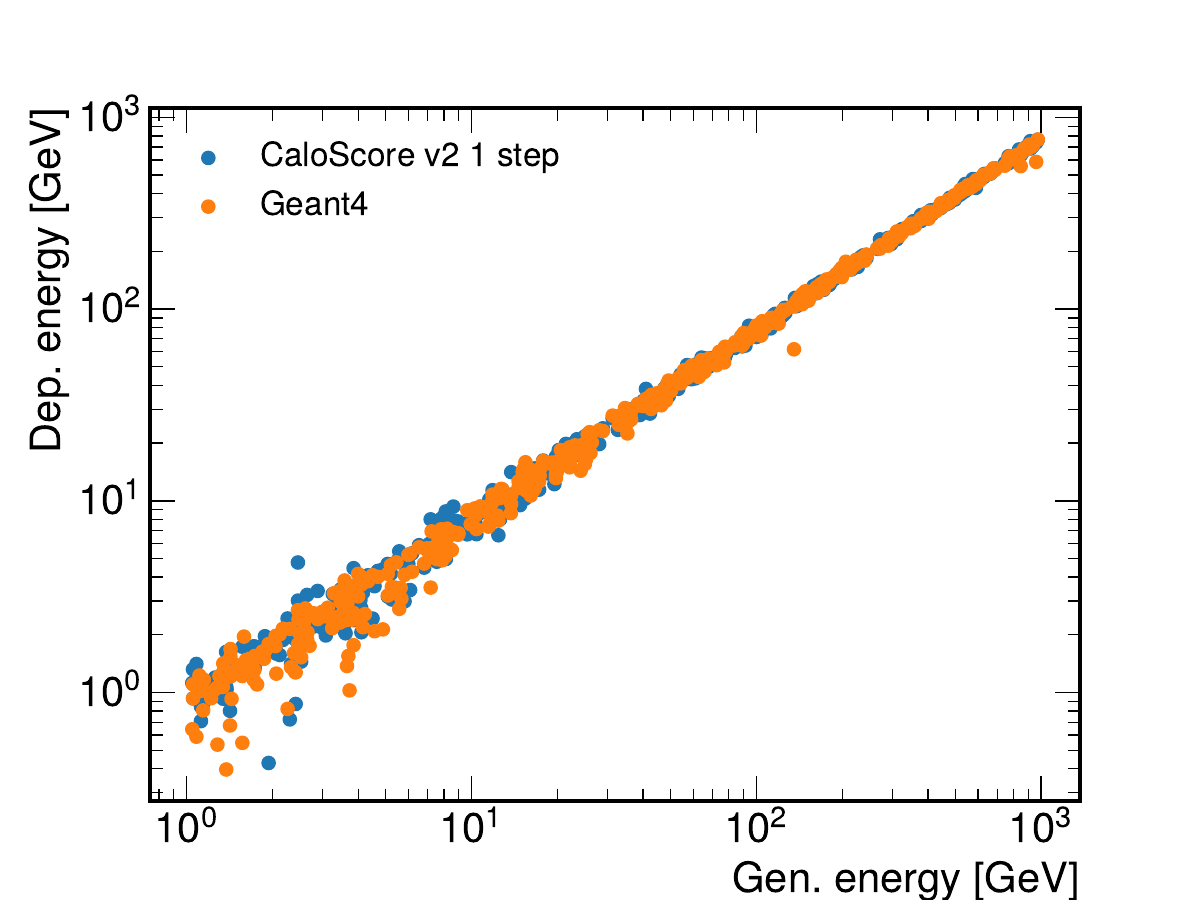}
    \includegraphics[width=0.31\textwidth]{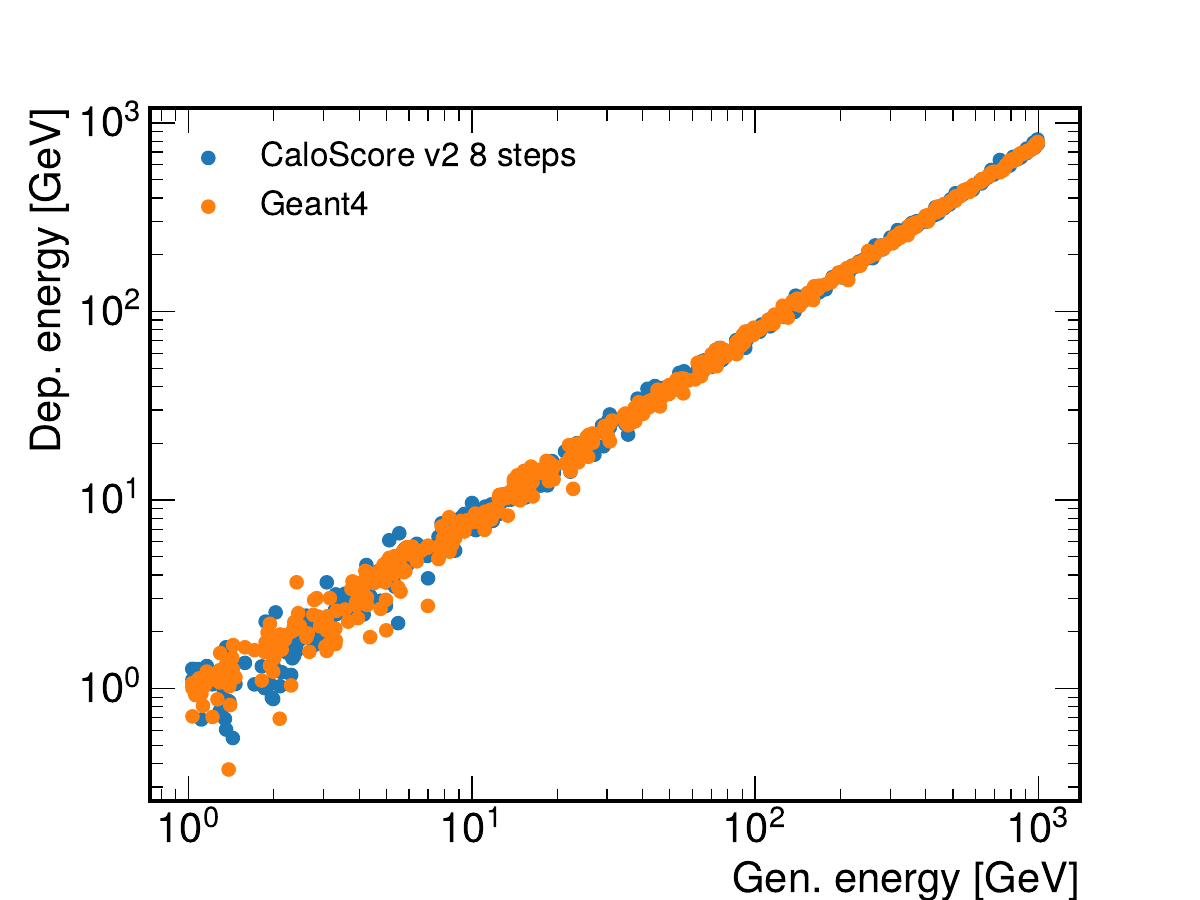}
    \includegraphics[width=0.31\textwidth]{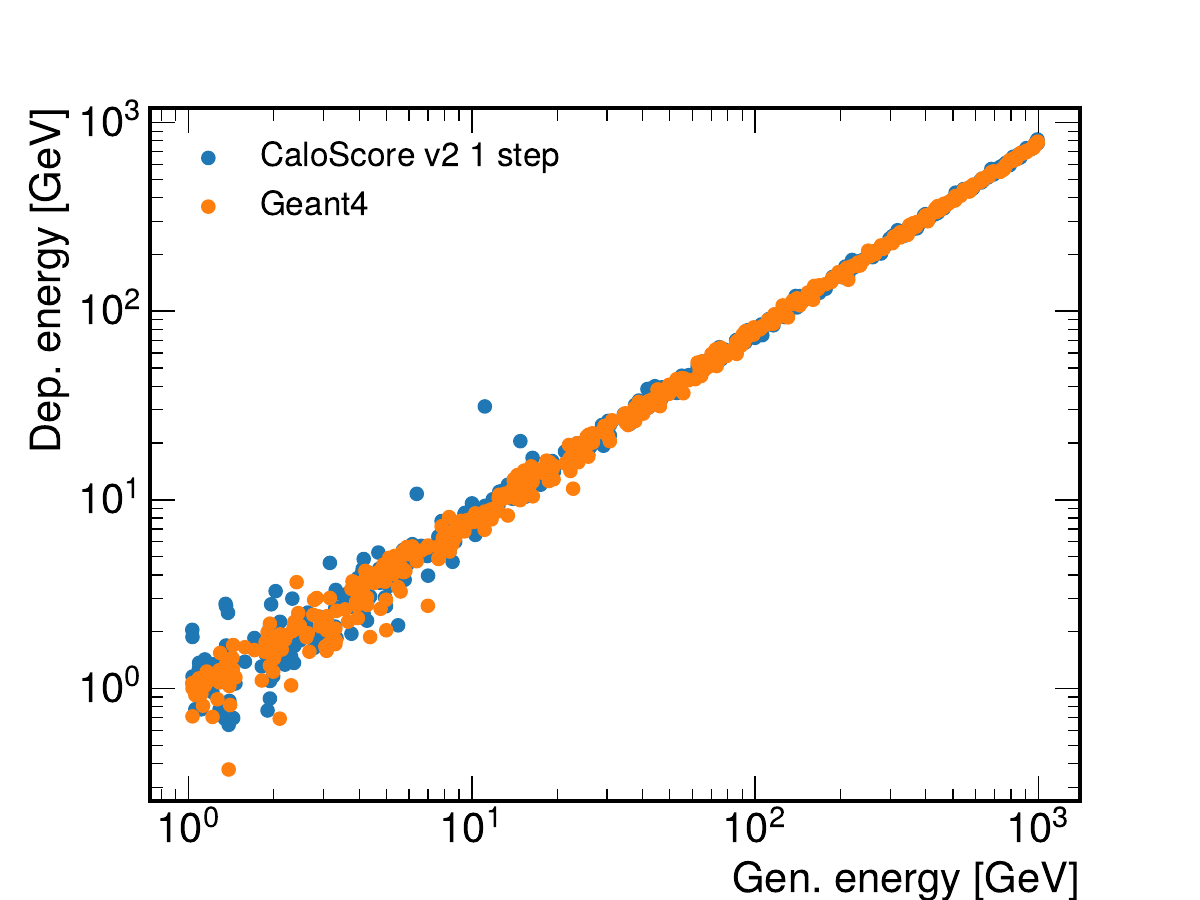}

\caption{Deposited energy versus generated energy in \geant~(orange) and \calosctwo~(blue) using different number of sampling steps. First row samples are generated using energies from dataset 1, second row from dataset 2 and third row from dataset 3.}
\label{fig:econd}
\end{figure*}

Finally, we investigate the energy conditioning of the model by comparing the distributions of the deposited energy and the energy of the incident particle. Results are presented in Fig.~\ref{fig:econd}.

In all cases the \calosctwo~model is able to correctly reproduce the deposited energy with spread similar to the one observed by the \geant~samples.

We also perform the so-called ``classifier test'' where a binary classifier is trained to distinguish generated samples from the samples produced by \geant, as used in the construction of GANs and proposed as a post-hoc metric in~\cite{caloflow1}. We use the official classifier and training schedule provided by the challenge to evaluate the results shown in Tables~\ref{tab:AUC_low} and~\ref{tab:AUC_high} for classifiers trained using either lower level inputs or high level distributions respectively. A similar classifier test was carried out when evaluating the original \calosc~performance, where an AUC of around 98\% was observed for all datasets. We should also point out that not only the classifier architecture, number of training epochs, and learning rates were different, but the initial preprocessing to convert datasets 2 and 3 to Cartesian coordinates makes the comparison of this metric deceptive.  On the other hand, \calosctwo~shows much lower values for both AUC and JSD metrics, showing a large improvement compared to the previous model. After distillation, we observe a degradation of the AUC and JSD in all datasets. Nevertheless, even the single-shot model observes AUC values significantly lower than 1 in all datasets. Direct comparisons with other models submitted to the Fast Calorimeter Challenge containing additional metrics will be made available in the forthcoming review paper.

\begin{table}[ht]
    \centering
	\small
    \caption{Area under the ROC curve (AUC) and Jensen-Shannon divergence (JSD) calculated based on the classifier trained using low level information.}
    \label{tab:AUC_low}
	\begin{tabular}{l|c|c|c|c|c|cc}
        Model &   \multicolumn{3}{c}{AUC/JSD}\\
        &  {\scriptsize dataset 1} & {\scriptsize dataset 2} & {\scriptsize dataset 3} \\
        \hline    
        
        \calosctwo & 0.758/0.155  & 0.597/0.023 & - \\ 
        \calosctwo~8 steps & 0.815/0.242  & 0.709/0.106 & 0.670/0.075\\
        \calosctwo~1 step& 0.878/0.367 & 0.755/0.157 & 0.6974/0.1002\\       
	\end{tabular}
\end{table}

\begin{table}[ht]
    \centering
	\small
    \caption{Area under the ROC curve (AUC) and Jensen-Shannon divergence (JSD) calculated based on the classifier trainined using high level information.}
    \label{tab:AUC_high}
	\begin{tabular}{l|c|c|c|c|c|cc}
        Model &   \multicolumn{3}{c}{AUC/JSD}\\
        &  {\scriptsize dataset 1} & {\scriptsize dataset 2} & {\scriptsize dataset 3} \\
        \hline        
        \calosctwo &  0.587/0.047 &  0.622/0.039&  -\\ 
        \calosctwo~8 steps& 0.6278/0.066  & 0.833/0.282 & 0.851/0.310 \\
        \calosctwo~1 step& 0.714/0.136& 0.846/0.305 & 0.880/0.376\\       
	\end{tabular}
\end{table}

We investigate the generation time required by \calosctwo~ and compare with previous results reported using the same hardware setup in Table.~\ref{tab:resources}. The baseline \calosctwo~ uses 512 time steps, value 5 times bigger than the original \calosc, leading to slower generation times. On the other hand, the distilled model with 8 time steps and the single-shot model decrease significantly the amount of time required, even compared to the initial \calosc~implementation and with a WGAN with similar overall architecture as \calosc. The reason for the reduction comes from \calosctwo~utilizing convolutional layers with smaller kernel sizes as opposed to \calosc. The larger kernel sizes were required to achieve a more precise result, since the transformation to cartesian coordinates increased the data sparsity, leading to a larger fraction of voxels without energy depositions.

\begin{table}[ht]
    \centering
	\small
    \caption{Number of dimensions, trainable parameter, and time to generate 100 new calorimeter showers for each dataset studied in this work. Generation times for \geant~are based on the average time required to generate samples over the energy range provided.}
    \label{tab:resources}
	\begin{tabular}{l|c|c|c|c|c|c|c|}
        Model &   \multicolumn{3}{c}{Time to 100 showers [s]}\\
        &  {\scriptsize dataset 1} & {\scriptsize dataset 2} & {\scriptsize dataset 3} \\
        \hline
        \calosc &   4.0 & 5.8 & 33.4 \\
        WGAN-GP &   1.3 & 1.33 & 2.06\\
        \geant &  $\mathcal{O}(10^2-10^3)$ & $\mathcal{O}(10^4)$  & $\mathcal{O}(10^4)$\\
        \calosctwo & 4.0 & 27.8 & 73.7\\ 
        \calosctwo~8 steps& 0.05 & 0.33& 1.71\\
        \calosctwo~ 1 step& 0.002& 0.010& 0.011\\       
	\end{tabular}
\end{table}

\section{Conclusions}
\label{sec:conclusions}

In this work we introduced \calosctwo~as a follow up to \calosc, a diffusion generative model for calorimeter shower simulation. Compared to its predecessor, \calosctwo~brings several changes to both increase the fidelity of the simulation and decrease the time required for the sampling of new observations. 

We evaluate the performance of \calosctwo~using the simulated samples created for the Fast Calorimeter Simulation Challenge 2022. We separate the generation process into two problems: generating the overall energy deposition in each layer of the calorimeter and generating the normalized voxel response. This modification improves the quality of the generated samples with better estimation of the overall energy deposition. Similarly, modifications to the network architecture through the use of attention layers increased the model performance without resulting in slower sampling times. Indeed, a single evaluation of \calosctwo~is now  faster compared to a single evaluation of \calosc. 

The sampling speed has also been reduced compared to \calosc~by a factor 500-2000 through the additional use of progressive distillation, a technique to iteratively reduce the number of time steps required during sampling. With this technique, we are able to reduce the generation to a single time step, resulting in the first single-shot diffusion models for detector simulation in collider physics. While the single-shot diffusion model shows a degradation in fidelity compared to the baseline \calosctwo, we still observe an overall good performance, also evidenced by the classifier test which is not able to distinguish samples from \geant~and~\calosctwo~ with perfect accuracy.  

Finally, progressive distillation shows that single-shot diffusion models can be achieved for fast and high fidelity simulation in collider physics. This observation motivates future work on reducing the performance degradation during the distillation process to retain the same level of precision as the initial diffusion model. Alternatively, smaller model architectures may be able to reduce even further the evaluation time by exploring additional symmetries present in the data. 

\section*{Code Availability}
The code used to produce all results presented in this paper are available at \url{https://github.com/ViniciusMikuni/CaloScoreV2}

\section*{Acknowledgments}

VM and BN are supported by the U.S. Department of Energy (DOE), Office of Science under contract DE-AC02-05CH11231. This research used resources of the National Energy Research Scientific Computing Center, a DOE Office of Science User Facility supported by the Office of Science of the U.S. Department of Energy under Contract No. DE-AC02-05CH11231 using NERSC award HEP-ERCAP0021099.

\bibliography{HEPML}
\bibliographystyle{apsrev4-1}

\clearpage
\appendix

\end{document}